\documentclass[conference]{IEEEtran}
\IEEEoverridecommandlockouts
\ifCLASSOPTIONcompsoc 
  \usepackage[nocompress]{cite}
\else  
  \usepackage{cite}
\fi
\usepackage{amsmath,amssymb,amsfonts}
\usepackage{stfloats}
\usepackage{algorithmic}
\usepackage{graphicx}
\usepackage{textcomp}
\usepackage{booktabs}
\usepackage{xcolor}
\usepackage{amsmath}

\DeclareMathOperator\erf{erf}
\usepackage{amssymb}
\usepackage{pgf,tikz}
\usepackage{mathrsfs}
\usetikzlibrary{arrows}
\usetikzlibrary{decorations.markings}
\usepackage{pgfplots}
\usepackage{pgfplotstable}
\usepackage{siunitx}
\usepackage{tikz}
\usepackage{rotating}
\usepackage{float}
\usepackage{multirow}
\pgfplotsset{compat=1.14}
\usepackage[caption=false]{subfig}
\usepackage{hyperref}
\usepackage{url}
\hypersetup{
    colorlinks=true,
    citecolor=black,
    linkcolor=black,
    filecolor=magenta,      
    urlcolor=black,
} 
\usepackage{algorithm}
\usepackage{algorithmic}
\usepackage[export]{adjustbox}
\usepackage{amsthm}
\usepackage{siunitx}
\usepackage[inline]{enumitem}

\newcommand{\de}[1]{\,\mathrm{d}#1}

\def\BibTeX{{\rm B\kern-.05em{\sc i\kern-.025em b}\kern-.08em
    T\kern-.1667em\lower.7ex\hbox{E}\kern-.125emX}}
\begin{document}

\title{SQLR: Short-Term Memory \textit{Q}-Learning\\ for Elastic Provisioning}
\author{\IEEEauthorblockN{Constantine Ayimba,~Paolo Casari,~Vincenzo Mancuso}%
    \thanks{Manuscript received xxxxxx xx, xxxx; \ldots. 
    The associate editor coordinating the review of this paper and approving it for publication was xxxxxxx. (\emph{Corresponding author: C.~Ayimba.)}
    }%
    \thanks{C. ~Ayimba (e-mail: constantine.ayimba@imdea.org) and V.~Mancuso (e-mail: vincenzo.mancuso@imdea.org) are with the IMDEA Networks Institute, 28918 Madrid, Spain.}%
    \thanks{C.~Ayimba is also with the University Carlos III of Madrid, 28918 Legan'{e}s, Madrid, Spain.}
    \thanks{P.~Casari is with the University of Trento, 38123 Povo (TN), Italy.}%
    \thanks{Digital Object Identifier XX.XXXX/XXXXXXXXXXX}%
    }
\maketitle

\begin{abstract}
As more and more application providers transition to the cloud and deliver their services on a  Software as a Service~(SaaS) basis, cloud providers need to make their provisioning systems agile enough to meet Service Level Agreements. At the same time they should guard against over-provisioning which limits their capacity to accommodate more tenants. To this end we propose SQLR, a dynamic provisioning system employing a customized model-free reinforcement learning algorithm that is capable of reusing contextual knowledge learned from one workload to optimize resource provisioning for other workload patterns. SQLR achieves results comparable to those where resources are unconstrained, with minimal overhead. Our experiments show that we can reduce the amount of provisioned resources by almost 25\% with less than 1\% overall service unavailability (due to blocking) while delivering similar response times as those of an over-provisioned system.
\end{abstract}
\section{Introduction}
There exists a growing tendency among Application Service Providers~(ASPs) to leverage cloud networks in delivering services to consumers. Such networks help reduce Capital Expenditure~(CAPEX) since ASPs do not have to deploy their own infrastructure. ASPs also reduce Operating Expenditure~(OPEX) as they only pay for the resources they use and do not incur network maintenance costs.

These factors mean that Cloud Service Providers~(CSPs) face increasing demands on finite resources. This is challenging given that ASPs have high expectations on performance\cite{Anselmi:2017}. CSPs must therefore balance the need for high Quality of Service~(QoS) guarantees with just the right amount of resources. In this way, cloud services can be delivered cost-effectively  without violating Service Level Objectives~(SLOs). Typical SLOs, stipulated in a Service Level Agreement~(SLA), are  service availability and response time.

SLO violations can have serious consequences for ASPs such as loss of users and revenue~\cite{Wang:2018}. In many cases, the CSP also has to pay penalties to the ASP~\cite{SLO_Penalty}. For CSPs, this creates the need for dynamic provisioning~(scaling) tools. Such tools increase the resources allocated to an application when sudden increases in demand occur (or are foreseen) and release them when they are not needed. These actions save costs for ASPs and free capacity for other CSP tenants.

Most state-of-the-art solutions to this problem leverage the continuous monitoring of application and system metrics to guide scaling decisions~\cite{ACM_Survey}. This means that any significant modification to the application necessitates a reconfiguration of the dynamic scaling system. The challenge therefore lies in minimizing the resources assigned to an application while guaranteeing service quality in the face of variable demand.

To address this challenge we have developed an application-agnostic system which leverages model-free reinforcement learning~(RL). Owing to the fact that workload profiles tend to be stochastic over short intervals, RL lends itself well to this problem, given its capability of learning from experience rather than training on static data-sets.

Unlike conventional RL our scaling scheme is not Markovian. It uses a modified \textit{Q}-Learning mechanism which keeps track of both the previous and current state. We define the state in terms of average resource utilization levels and the number of active VMs. This modification makes it possible for our system to learn different policies for different traffic profiles by implicitly leveraging the rate of change in utilization levels as a distinguishing parameter between profiles.

Concretely, our main contributions are:
\begin{enumerate}
\item A flexible scaling agent that
\begin{itemize}
    \item is horizontal, i.e., adapts the number of virtual machines (VMs) allocated to a service, not their resources (e.g. virtual CPUs or memory);
    \item given high-level objectives, learns the optimal trade-off between the accepted level of service availability and the corresponding resource costs, and provisions accordingly even in the presence of challenging workloads;
    \item adapts to different workloads by learning policies based on their resource utilization patterns and not on the workloads \textit{per se};
    \item progressively improves with every scaling decision, resulting in better performance with regard to service reliability and availability.
\end{itemize}
\item A configuration-agnostic Admission Control Virtual Network Function~(VNF) based on \textit{Q}-Learning, that learns the most suitable action to take given the level of resource utilization reported by a Virtual Machine (VM) instance.
\item A weighted fair learning mechanism that encourages exploration in unfamiliar states~and exploitation for better known states; this increases the likelihood of selecting optimal actions prior to full convergence, when the system is still acting according to partially developed policies.
\end{enumerate}

The rest of this article is organized as follows: Section~\ref{relatio} examines prior approaches to scaling of cloud resources, while in Section~\ref{apologia} we discuss the adaptations we made to conventional \textit{Q}-Learning to achieve a short-term memory algorithm capable of learning multiple policies. In Section~\ref{experimentum}, we outline the experiments made to test the scaling and admission control algorithms. We discuss the results of our approach compared to other methods in Section~\ref{resultet}. In Section~\ref{FinisTerrae}, we finally provide the main conclusions and discuss future directions along this line of research.
\section{Related Work}
\label{relatio}
The most commonly used approaches in making automated scaling decisions are rule-based schemes. These approaches rely on leveraging resource utilization thresholds which can be fixed such as in commercial tools as Rightscale~\cite{Rightscale} and Amazon\textquoteright s EC2~\cite{EC2}, or loosely defined as the fuzzy logic variants proposed by \cite{Farokhi}, \cite{Hasan2012}, \cite{Jamshidi} among others. These methods require sufficient knowledge of the cloud application in order to define the operating bounds correctly.

The authors of~\cite{Math_Q} propose a theoretical, model based, RL approach to cloud resource allocation which factors in both net gains for the CSP and SLO violations. Their model, however, assumes high predictability in arrival rates and system responses, both of which are highly stochastic.

In~\cite{BibalBenifa2019}, the authors propose an on-policy RL horizontal scaling system. Their technique leverages a one-step temporal difference scheme with multiple coordinating agents. Application specific targets of throughput and response time are used as inputs to reward functions. However, this method requires to approximate the action-values, and is thus susceptible to biasing action selection. The latter characteristic makes it ill-suited to highly dynamic workloads.

The authors of~\cite{iBaloon} implement a vertical scaling engine based on \textit{Q}-learning. Distributed RL agents adjust the CPU, memory and bandwidth allocations to a set of active VMs handling different applications. The effect on the applications in terms of response time and throughput act as inputs for the rewards fed back to the agent. Application agents maintain fine-grained SLA metrics for each application.

An RL based agent is proposed in~\cite{RL_Migration} which triggers the migration of VMs from under-utilised servers, which can then be powered off. Utilization bounds that trigger decisions in this scheme are predetermined. This implies that the response of the agent will be compromised, should the system configuration change in a significant way. 

The authors of~\cite{Gandhi} use a combination of a queue model and Extended Kalman Filtering~(EKF) to carry out horizontal scaling, i.e., the addition or removal of VMs to a resource pool providing a given service. They use a 3-tier cloud application with 3 classes of requests to generate the measurement model. The model, enhanced by EKF, estimates the response times given the workload as input. These estimates trigger an appropriate horizontal scaling operation.

Other methods seek to characterize the workload and make resource allocations accordingly. In~\cite{Vasic}, Vasic \textit{et al}. classify workloads based on recurring patterns. Optimized resource allocations for these patterns are derived and re-used every time the same patterns are detected in a new workload.

Ibidunmoye \textit{et al}.~\cite{Ibidunmoye} use a modified \textit{Q}-Learning scheme in order to carry out vertical scaling~(the addition of system resources e.g. virtual CPUs on VMs while they are running). Their state space is based on a fuzzy logic combination of response times and utilization levels. They employ several cooperating agents to simultaneously explore the state space in order to speed up convergence to a given policy.

The authors of~\cite{Aggressive} propose a scheme based on \textit{Q}-Learning and heuristics that immediately over-provisions resources when an increase in the workload is detected. It then gradually de-allocates extra resources. The objective of this scheme is to reduce SLO violations that occur when the workload increases suddenly but resources are added conservatively.

In~\cite{AWAHCI}, the authors propose a system which profiles resource capacities, predicts the subsequent workload pattern over a monitoring window and scales the system accordingly based on a trade-off between scaling costs and SLOs.

The authors of~\cite{iSpot} develop an LSTM algorithm to offload big data analytics workloads to Amazon EC2 spot instances. The scheme trades-off the very cheap prices of these transient servers with their unreliability due to revocation. Given their focus on ASP cost objectives, this work is largely orthogonal to our approach which considers costs incurred by the CSPs.

In~\cite{Leitner}, the provisioning problem is considered in terms of costs for the provider. The authors propose a system that schedules resources in a bid to minimize the costs incurred due to SLO violations and those resulting from leasing cloud resources. This work assumes that such costs are known well in advance, and that Billing Time Units~(BTUs) for leasing resources are both coarse-grained~(in the order of hours) and fixed. However, recent proposals on cloud brokerage~\cite{BTU} promise greater flexibility by making BTUs much more fine grained, providing better cost-effectiveness for tenants with short-lived requirements.

In summary, whereas rule-based schemes are simple and easy to implement, setting the correct thresholds requires specialized cloud application domain knowledge and awareness of resource configurations. Other state-of-the-art scaling methods require that the service response times, the workload, or both are continuously monitored and measured. In many cases, obtaining such data requires real-time analysis of logs, which may lead to significant overhead in large-scale systems. Workloads, in particular, may also exhibit unpredictable behavior~\cite{Wang2012} resulting in premature scaling directives.

Our scheme, instead, infers the changes in workloads by monitoring how the system responds to them. This provides a more robust basis for decision-making, even in large deployments. Different workloads trigger distinct state transition sequences, resulting in new policies being learned, in addition to those already learned. Given this, the system will scale for future workloads that exhibit combinations of the already observed patterns, without need for further training.
\section{SQLR Design}
\label{apologia}
\subsection{Problem Statement}
The problem we tackle in this paper can be stated as follows: \emph{Minimize the number of instantiated VMs that run a given service, under the constraint that the probability to block a service request remains below a predefined threshold. Maximize the number of jobs served by horizontally scaling the number of VMs as workload evolves over time.}

\begin{figure}[h]
    \centering
    \includegraphics{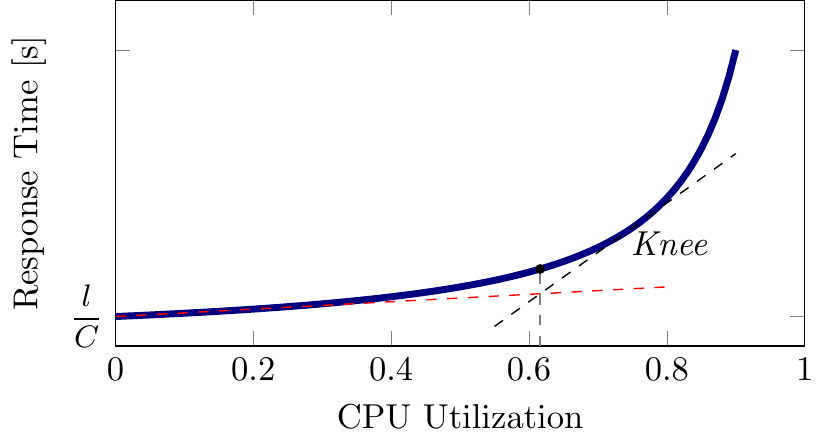}
    \caption{Response time variation with load based on the queuing theory analysis presented in~\cite{Kleinrock:67}. The variation is approximately linear just below the \textquotedblleft \textit{knee}.\textquotedblright\ If utilization levels are kept below this value, response times are highly likely to be predictable and reliable.}
    \label{fig:Resp_CPU}
\end{figure}

To formulate the problem, let $X_{ji}(t)=1$ if job $j$ arrives at time $t$ and is assigned to VM $i$, and $0$ otherwise. Also, let $Y_{ji}(t) = 1$ if job $j$ is running on VM $i$ at time $t$, and $0$ otherwise.
Call $V(t)$ the number of VMs activated at time $t$, $V_{\max}$ the maximum number of VMs reserved for an ASP, and $A(t)$ the set of jobs arriving at time $t$. Let $P_{\mathrm{blk}}$ be the ideal blocking probability set out in the SLA, and $\rho_j$ be a function that determines the contribution of job $j$ to the utilization level of a given VM. Finally, let $x_{\mathrm{bnd}}$ be the utilization level above which response times become unpredictable~(cf. Fig.~\ref{fig:Resp_CPU}).

This can be expressed formally as:
\begin{subequations}
    \label{eq:Optime}
    \begin{align}
        \min \; & \frac{1}{T}\int_0^T V(t) \de{t} \label{eq:Optime.obj} \\        
        \text{s.t.} \; & \dfrac {\int_0^T  \sum_{i\in V(t)}\sum_{j\in A(t)}X_{ji}(t)\de{t}}{\int_0^T A(t)\de{t}} \ge 1-P_{\text{blk}}  \label{eq:Optime.c1} \\
        &  1 \le V(t) \le V_{\max}\ \ \ \ \ \ \forall t, \label{eq:Optime.c2} \\
          &\sum_{i\in V(t)} X_{ji}(t) \le 1 , \label{eq:Optime.c4} \\
          &\sum_{i\in V(t)} Y_{ji}(t) \le 1 \label{eq:Optime.c5} \\
          &\sum_{j\in J(t)} \rho_j Y_{ji}(t) \le x_{\mathrm{bnd}} \; \forall i \in \{1,\ldots, V(t)\}, \label{eq:Optime.c3}
    \end{align}
\end{subequations}
Constraint~\eqref{eq:Optime.c1} ensures that the number of jobs dropped are kept within SLA bounds for service unavailability. Constraint~\eqref{eq:Optime.c2} ensures that the number of VMs reserved for an ASP are bounded. Constraint~\eqref{eq:Optime.c4} specifies that a job can only be assigned to one VM, and~\eqref{eq:Optime.c5} indicates that a given job can only be running on one VM at a time.

Constraint~\eqref{eq:Optime.c3} ensures that jobs admitted to a VM will not suffer unpredictable responses by driving the utilization of the VM above an allowable level. An illustration of this phenomenon, arising from the extensive analysis presented in~\cite{Kleinrock:67}, is shown in Fig.~\ref{fig:Resp_CPU}. 
We remark that, to meet constraint~\eqref{eq:Optime.c3}, we need an admission controller capable of adapting to the system configuration. We also need to employ load balancing to even out utilization levels across all active VMs.

Besides $A(t)$ and $\rho_j$ being unknown functions, the problem presented in~\eqref{eq:Optime.obj} is a variant of the knapsack problem and can only be solved approximately in polynomial time.

To solve \eqref{eq:Optime.obj} and implement appropriate admission control, we elect to use \textit{Q}-Learning, given its versatility in finding near optimal solutions in uncertain settings~\cite{Sutton:2018}.

\begin{table*}[b!]
\centering
\caption{Key notation employed in the definition of SQLR}
\textbf{\label{tbl:notations}}
\renewcommand{\arraystretch}{1.15}
\renewcommand{\baselinestretch}{1.00}\footnotesize
\begin{tabular}{m{0.06\linewidth} m{0.19\linewidth} m{0.67\linewidth} @{\hspace{1mm}}}
\midrule
\textbf{Variable}      & \textbf{Meaning}            & \textbf{Description}                                                                                                           \\ 
\midrule
$x_{\mathrm{bnd}}$   & Utilization  upper bound   & The utilization level above which response times become unpredictable, inferred from~\cite{Kleinrock:67} as 62\%                              \\ 
$x_{\mathrm{tgt}}$  &  Utilization target   & The level  on which the admission controller is trained. Set as 60\% to ensure a safety margin from $x_{\mathrm{bnd}}$                            \\ 
$x_{\mathrm{lim}}$  &  Utilization admission limit   & The practical limit of resource utilization obtained after training the admission controller                          \\ 
$\theta$       & Resource cost modifier           & Multiplier that weighs the cost of deploying resources in the reward function                                         \\ 
$\beta$      & Blocking probability   modifier           & Multiplier that weighs the blocking rate in the reward function                                                \\ 
$\epsilon_{\mathrm{min}}$ & Minimum randomness factor  & The minimum probability of selecting an off-policy action after convergence. We set this at 0.         \\ 
$\gamma$       & Discount rate in $(0,1]$         & Expresses the current value of a future reward due to the present action. We set $\gamma = 0.8$.                        \\ 
$P_{\mathrm{blk}}$  &  Target blocking probability            & We set $P_{\mathrm{blk}}=0.001$, corresponding to service availability of 99.9\% \\ 
$R_{\mathrm{min}}$  & Minimum reward      & A small, positive reward accrued when the scaling agent maintains the blocking probability within $P_{\mathrm{blk}}$ \\
\midrule
\end{tabular}
\end{table*}

\subsection{System Design}
\label{SysDes}
\begin{figure}[b]
    \centering
    \begin{tikzpicture}[line cap=round,line join=round,>=triangle 45,x=1.0cm,y=1.0cm,scale=2.5,every text node part/.style={align=center}]
\draw [line width=0.5pt] (-2.4,0.)-- (-1.4,0.);
\draw [line width=0.5pt] (-2.4,-0.4)-- (-1.4,-0.4);
\draw [line width=0.5pt] (-2.4,-0.6)-- (-1.4,-0.6);
\draw [line width=0.5pt] (-2.4,-1.)-- (-1.4,-1.);
\draw [line width=0.5pt] (-2.4,-1.2)-- (-1.4,-1.2);
\draw [line width=0.5pt] (-2.4,-1.6)-- (-1.4,-1.6);
\draw [line width=0.5pt] (-2.4,-1.8)-- (-1.4,-1.8);
\draw [line width=0.5pt] (-2.4,-2.2)-- (-1.4,-2.2);
\draw [line width=0.5pt] (-2.4,-1.8)-- (-2.4,-2.2);
\draw [line width=0.5pt] (-1.4,-1.8)-- (-1.4,-2.2);
\draw [line width=0.5pt] (-2.4,0.)-- (-2.4,-0.4);
\draw [line width=0.5pt] (-1.4,0.)-- (-1.4,-0.4);
\draw [line width=0.5pt] (-2.4,-0.6)-- (-2.4,-1.);
\draw [line width=0.5pt] (-1.4,-0.6)-- (-1.4,-1.);
\draw [line width=0.5pt] (-2.4,-1.2)-- (-2.4,-1.6);
\draw [line width=0.5pt] (-1.4,-1.2)-- (-1.4,-1.6);
\draw [line width=0.5pt] (-2.6,0.2)-- (-1.2,0.2);
\draw [line width=0.5pt] (-3.4,-0.8)-- (-3.4,-1.2);
\draw [line width=0.5pt] (-3.4,-1.2)-- (-3.,-1.2);
\draw [line width=0.5pt] (-3.,-0.8)-- (-3.,-1.2);
\draw [line width=0.5pt] (-3.4,-0.8)-- (-3.,-0.8);
\draw [line width=0.5pt] (-3.4,-0.4)-- (-3.,-0.4);
\draw [line width=0.5pt] (-3.4,0.)-- (-3.4,-0.4);
\draw [line width=0.5pt] (-3.4,0.)-- (-3.,0.);
\draw [line width=0.5pt] (-3.,0.)-- (-3.,-0.4);
\draw [line width=0.5pt] (-0.8,0.)-- (-0.8,-1.6);
\draw [line width=0.5pt] (-0.8,-1.6)-- (0.,-1.6);
\draw [line width=0.5pt] (0.,0.)-- (0.,-1.6);
\draw [line width=0.5pt] (-0.8,0.)-- (0.,0.);
\draw [line width=0.3pt,dash pattern=on 2.0pt off 2.0pt] (-0.6,-0.2)-- (-0.2,-0.2);
\draw [line width=0.3pt,dash pattern=on 2.0pt off 2.0pt] (-0.6,-0.4)-- (-0.2,-0.4);
\draw [line width=0.3pt,dash pattern=on 2.0pt off 2.0pt] (-0.6,-0.6)-- (-0.2,-0.6);
\draw [line width=0.3pt,dash pattern=on 2.0pt off 2.0pt] (-0.6,-0.8)-- (-0.2,-0.8);
\draw [line width=0.3pt,dash pattern=on 2.0pt off 2.0pt] (-0.6,-0.6)-- (-0.6,-0.8);
\draw [line width=0.3pt,dash pattern=on 2.0pt off 2.0pt] (-0.2,-0.6)-- (-0.2,-0.8);
\draw [line width=0.3pt,dash pattern=on 2.0pt off 2.0pt] (-0.6,-0.2)-- (-0.6,-0.4);
\draw [line width=0.3pt,dash pattern=on 2.0pt off 2.0pt] (-0.2,-0.2)-- (-0.2,-0.4);
\draw [line width=0.3pt,dash pattern=on 2.0pt off 2.0pt] (-0.6,-1.2)-- (-0.6,-1.4);
\draw [line width=0.3pt,dash pattern=on 2.0pt off 2.0pt] (-0.6,-1.4)-- (-0.2,-1.4);
\draw [line width=0.3pt,dash pattern=on 2.0pt off 2.0pt] (-0.6,-1.2)-- (-0.2,-1.2);
\draw [line width=0.3pt,dash pattern=on 2.0pt off 2.0pt] (-0.2,-1.2)-- (-0.2,-1.4);
\draw [line width=0.5pt] (-2.6,0.2)-- (-2.6,-2.4);
\draw [line width=0.5pt] (-1.2,0.2)-- (-1.2,-2.4);
\draw [line width=0.5pt] (-2.6,-2.4)-- (-1.2,-2.4);
\draw [->,line width=0.1pt] (-3.2,0.4) -- (-3.2,0.);
\draw [->,line width=0.1pt] (-0.4,0.4) -- (-0.4,0.);
\draw [->,line width=0.1pt] (-2.8,-0.2) -- (-2.4,-0.2);
\draw [->,line width=0.1pt] (-3.2,-0.4) -- (-3.2,-0.8);
\draw [->,line width=0.1pt] (-3.2,-0.8) -- (-3.2,-0.4);
\draw [line width=0.5pt] (-3.,-1.)-- (-2.8,-1.);
\draw [line width=0.5pt] (-2.8,-1.)-- (-2.8,-0.2);
\draw [->,line width=0.1pt] (-0.8,-0.2) -- (-1.4,-0.2);
\draw [->,line width=0.1pt] (-1.9,-0.4) -- (-1.9,-0.6);
\draw [->,line width=0.1pt] (-1.9,-1.) -- (-1.9,-1.2);
\draw [->,line width=0.1pt] (-1.9,-1.6) -- (-1.9,-1.8);
\draw [->,line width=0.1pt] (-2.5,-0.8) -- (-2.4,-0.8);
\draw [line width=0.5pt] (-2.4,-2.)-- (-2.5,-2.);
\draw [line width=0.5pt] (-2.5,-2.)-- (-2.5,-0.8);
\draw [line width=0.5pt] (-3.2,0.4)-- (-0.4,0.4);
\draw [->,line width=0.1pt] (-1.4,-1.4) -- (-0.8,-1.4);
\draw (-3.35,-0.12) node[anchor=north west] {LB};
\draw (-3.35,-0.92) node[anchor=north west] {AC};
\draw (-0.6,-0.2) node[anchor=north west] {VM$_1$};
\draw (-0.6,-0.6) node[anchor=north west] {VM$_2$};
\draw (-0.6,-1.2) node[anchor=north west] {VM$_n$};
\draw (-0.79,-0.0) node[anchor=north west] {Hypervisor};
\draw (-2.4,0.2) node[anchor=north west] {Scaling Engine};
\draw (-2.2,-0.1) node[anchor=north west] {Monitor};
\draw (-2.2,-0.61) node[anchor=north west] {Action \\ Analyzer};
\draw (-2.1,-1.3) node[anchor=north west] {Scaler};
\draw (-2.3,-1.81) node[anchor=north west] {Action \\ Evaluation};
\draw [line width=0.5pt] (-3.5,0.5)-- (-3.5,-2.5);
\draw [line width=0.5pt] (-3.5,-2.5)-- (-1.,-2.5);
\draw [line width=0.5pt] (-3.5,0.5)-- (-1.,0.5);
\draw [line width=0.5pt] (-1.,0.5)-- (-1.,-2.5);
\draw (-3.37,-1.7) node[anchor=north west] {SQLR};
\begin{scriptsize}
\draw [fill=black] (-0.4,-0.9) circle (0.2pt);
\draw [fill=black] (-0.4,-1.) circle (0.2pt);
\draw [fill=black] (-0.4,-1.1) circle (0.2pt);
\end{scriptsize}
\end{tikzpicture}
    \caption{SQLR block diagram. \textquotedblleft LB\textquotedblright\ is the Load Balancer VNF and \textquotedblleft AC\textquotedblright\ is the Admission Control VNF.}
    \label{fig:SQLR_Block}
\end{figure}
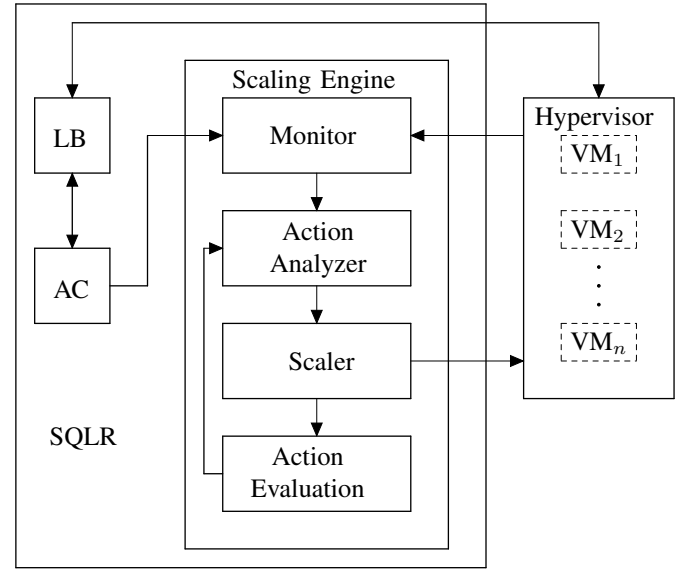
A block diagram of the provisioning system we have developed~(named SQLR in the following and read as \textquotedblleft scaler\textquoteright) is shown in Fig.~\ref{fig:SQLR_Block}. It comprises: a Load Balancer~(LB) VNF, an Admission Control~(AC) VNF and a scaling engine. The latter leverages model-free reinforcement learning with inputs from system monitors in order to determine the appropriate horizontal scaling action. It then issues directives to the hypervisor to instantiate or shut down the VMs that deliver a cloud application as a service.

In a manner similar to the approach taken by~\cite{Math_Q}, we treat admission control and horizontal scaling as sequential decisions. Such decisions can be formalised as Markov Decision Processes~(MDPs) with a set of actions \textit{A}, states \textit{S}, state transition probabilities \textit{T} and a reward function \textit{R}.
\begin{equation}
    \{T,S,A,R\}.
    \label{eq:mdp}
\end{equation}

We consider the admission control and scaling mechanism as agents capable of epochal or episodic reinforcement learning. An episode or epoch consists of an agent observing its state, taking a given action permissible in that state, monitoring the environment to compute and accrue the reward value corresponding to that action, and transitioning to a new state.

For an admission control agent, the permissible actions are to ADMIT or DROP a request. An ADMIT or DROP action results in a VM transitioning from one level of utilization to another with some probability. By defining the state as the utilization level of the VM handling the request, we are able to formulate this as a RL problem. We structure the reward values based on how stable the resulting state is with respect to service times (inferred from the utilization level, cf. Fig.~\ref{fig:Resp_CPU}).

For the horizontal scaling agent, the actions are increasing, decreasing or maintaining the number of VMs. The state, in this case, is defined by three values: the average system-wide utilization over the previous epoch; the average system-wide utilization in the current epoch; and the number of active VMs. Therefore, the actions taken by the scaling agent results in a change in this set of values, which can be considered a transition in system state. We structure the reward function to include both the resulting blocking rate and the consequent number of VMs used.

The key notations used in describing our system is given in Table~\ref{tbl:notations}.

By tracking how much reward an action receives and obtaining the state transition probabilities, the optimal actions (that would yield the highest accumulated reward given a particular system state) could be determined, and the agent programmed to carry them out. However, a solution to the MDP is impractical given that the transition probabilities can vary widely depending on the workload and configuration of the system. A practical way that has been applied in many similar cases, where an exact solution to the MDP is intractable, is \textit{Q}-Learning \cite{Watkins89}. In this method, the agent develops a mapping of states to actions (known as the \textit{Q}~function) by tracking the accumulated reward or (\textquotedblleft\textit{Q}-value\textquotedblright) for each state-action pair.

With reference to Table~\ref{tbl:notations}, which summarizes the key notations used in this paper, we now explain the design principles and behaviour of the scaling and admission control agents.
At epoch $t$ the optimal action-value function $q^*$ is approximated as:~\cite{Sutton:2018}:
\begin{equation}
    Q(S^{(t)},A^{(t)}) \leftarrow \alpha  R' + (1-\alpha) Q(S^{(t)}, A^{(t)}),
    \label{eq:Qoptimi}
\end{equation}\\
where 
\begin{equation}
    R' = R^{(t+1)} + \gamma \max_{\mathrm{a}}{Q(S^{(t+1)}, a)},
\end{equation}
$Q(S^{(t)},A^{(t)})$ is the action-value, and $R^{(t+1)}$ is the immediate reward the agent receives after taking action $a$ and ending up in state $S^{(t+1)}$,  whose action-value is $Q(S^{(t+1)}, a)$. The fraction $\gamma$ anticipates the contribution that future rewards will make towards the immediate one \cite{Watkins89}.

The use of a fixed learning rate $\alpha$ in \eqref{eq:Qoptimi} assumes that all states are visited evenly during training~\cite{Sutton:2018}. Depending on the formulation of the state space, this may not always be the case. Further, the update process given by \eqref{eq:Qoptimi} typically leads to a stochastic policy with values oscillating slightly about an estimated expected value. To ameliorate this effect, a modified reward mechanism can be used which takes into account the number of times the given state is visited. This method follows closely the algorithm for the online computation of the mean given in equation \eqref{eq:modified_mean}: 
\begin{equation}    
    \mu_{n} = \dfrac{1}{n}\sum_{k=1}^{n}\left(X_{k}\right) = \dfrac{1}{n} \left(X_{n} + (n-1)\mu_{n-1} \right).    
    \label{eq:modified_mean}
\end{equation}
The \textit{Q} function update then becomes:
\begin{equation}
    Q(S^{(t)},A^{(t)})  \leftarrow \dfrac{1}{n} \left[ \Delta + (n-1)Q(S^{(t)}, A^{(t)}) \right],
    \label{eq:modified_Q}
\end{equation}\\
where 
\begin{equation}
    \Delta = R' - Q(S^{(t)}, A^{(t)}) ,
\end{equation}
and $n$ is the number of episodes (prior to the current action) that the agent acted with $(S^{(t)},A^{(t)})$ as the preceding state-action pair.

We modify the update mechanism by using the discounted reward $\Delta$ instead of the immediate reward $R'$ in \eqref{eq:modified_Q}. This helps reduce the chances of wrongly estimating the mean action value at the initial learning phases. The modified update outlined in \eqref{eq:modified_Q} also guarantees that a stationary policy will eventually be developed: in fact, the update value on the right hand side of \eqref{eq:modified_Q} becomes progressively smaller as the number of episodes increases. 

The agent is trained by initially encouraging random actions (exploration phase). As it develops a policy, it progressively acts less randomly by choosing the actions yielding the highest reward at a given system state (exploitation phase). This is accomplished by employing $\epsilon$-greedy action selection~\cite{Sutton:2018}. In this scheme, the action that yields the highest reward is chosen with a probability $(1-\epsilon)$ and a random action chosen with a probability $\epsilon$.

We remark that not all states are visited with the same frequency. Therefore, a global assignment and decrease of $\epsilon$ may bias the learned policy towards the most visited states. To avoid this, we employ a scheme that reduces $\epsilon$ independently for each state depending on the number of times $i$ that such state is visited. This accelerates the learning process and performance by encouraging exploration for the least visited states, while exploiting optimal actions for the most visited states. Specifically, se set 
\begin{equation}
    \epsilon = \begin{cases}
    1-\dfrac{i}{M},& \text{if } i<M\\
    \epsilon_{\mathrm{min}},& \text{if } i\geqslant M,
\end{cases}
    \label{eq:epsiloner}
\end{equation}
where $M$ is a design parameter representing the number of statistically significant visits that should result in convergence to a stable policy. A state is considered to have achieved convergence when its associated $\epsilon$ is equal to $\epsilon_{\mathrm{min}}$.

In order for the system to perform satisfactorily even before it has fully converged, we devise a weighted fair guided exploration scheme. In this scheme, at learning instance $i$, the probability $P^{(i)}$ of selecting an action $a$ depends on its present action value $Q^{(i)}(s,a)$ and on the number of times $n^{(i)}$ that it has previously been selected when the system was in state $s$:

\begin{equation}
    P^{(i)}  = \begin{cases}
    \dfrac{1}{L},& \text{for } \Psi^{(i)}=0\\
    \dfrac{\Psi^{(i)}(1-\tanh{\phi^{(i)}})}{\sum_{k=1}^{L}\Psi^{(i)}_{k}(1-\tanh{\phi^{(i)}_{k}})}, & \text{for } \Psi^{(i)}>0
    \end{cases}
    \label{eq:selectioner}
\end{equation}\\
where, 
\begin{equation}
    \begin{split}
    \Psi^{(i)} & = Q^{(i)}(s,a) +\sum_{j=1}^{L}|Q^{(i)}_{j}(s,a)|, \\
    \phi^{(i)} & = \dfrac{n^{(i)}}{i},
    \end{split}
\end{equation}
and $L$ is the number of permissible actions in state $s$. Note that $\Psi^{(i)} \geqslant 0$ is used in place of the action value $Q^{(i)}(s,a)$ which, if negative, would result in unfeasible probabilities. The hyperbolic tangent is a suitable weighting function, as $0\leqslant \tanh(\phi)\leqslant 1$ for $\phi\geqslant0$.

With the above strategy, a workable compromise between exploration and exploitation is achieved, curtailing the detrimental effects of unguided exploration on performance.

\subsection{Load Balancer}
Given that this component does not constitute a contribution of our work, we only mention it briefly here. Our Load Balancer works by choosing the VM with the lowest, most recently logged CPU utilization at the time a request is received. This method is reminiscent of server state based strategies used for classic web traffic~\cite{Load_Balancing}. CPU utilization is logged at \si{1\second} intervals. This policy gives a high probability that the available resources will be evenly loaded. In this way, the disparity in overall response times is also reduced.
\subsection{Admission Control}
In a resource-constrained system, an admission control mechanism ensures that the system does not take on more tasks than it can satisfactorily handle. According to \cite{Kleinrock:67}, it is possible to use the theoretical utilization bound ($x_{\mathrm{bnd}}$) to make the admission decision. However to obtain the best results for an actual system it is necessary to learn an appropriate admission limit ($x_\mathrm{lim}$) that takes the system configuration into account. As mentioned in Section~\ref{SysDes}, we do this by treating admission control as a sequential decision process. The action space in this case is defined by the mutually exclusive options: 
\begin{enumerate}
  \item ADMIT the request;
  \item DROP the request.
\end{enumerate}

The state space, however, is derived from the quantized levels of resource utilization on the VM serving the request. In this work, the resource considered is CPU utilization, as it is a low-level metric that correlates well with the workload,  and does not require any domain-specific knowledge of the deployed application~\cite{Aggressive}. Bearing in mind that response times are greatly impacted by how busy the system is, the upper threshold on utilization is chosen as the one beyond which the service times will likely violate the agreed SLO. This threshold is used as a target to determine the rewards/penalties the admission controller will accrue as it builds a policy using \textit{Q}-Learning. 

The CPU utilization threshold is chosen based on the analytical results relating response times to occupancy in a processor sharing queue described in~\cite{Kleinrock:67}. The time $T$, taken by a processor with capacity $C$ operations per second to service a request requiring $\ell$ operations is be given by: 
\begin{equation}
    T(\rho)  = \dfrac{\ell}{C-\rho}, \qquad
    \rho  =\dfrac{\lambda}{\mu},
    \label{eq:QService}
\end{equation}
where $\lambda$, is the arrival rate~(workload) and $\mu$ is the departure rate. The occupancy of the processor $\rho$ is here considered as its utilization level.

\begin{figure}
    \centering    
    \begin{minipage}{\linewidth}
    \usepgfplotslibrary{fillbetween}
\begin{tikzpicture}
\pgfplotsset{every axis y label/.append style={font=\color{black!50!red}}}
\begin{axis}[%
scale=0.2,
width=12.917in,
height=6.647in,
at={(2.167in,0.897in)},
scale only axis,
every outer x axis line/.append style={black},
xticklabels={40,45,50,55,60,65,70,75,80},
xtick={40,45,50,55,60,65,70,75,80},
xmin=40,
xmax=80,
xlabel={Reference Time [s]},
every outer y axis line/.append style={black!50!red},
every y tick label/.append style={font=\color{black!50!red}},
yticklabels={0,0.3,0.6,0.9,1.2,1.5,1.8,2.1,2.4},
ymin=-2.5,
ymax=1.5,
ytick={-2.5,-2.0,-1.5,-1.0,-0.5,0,0.5,1.0,1.5},
ylabel={Service Time [s]},
]
\addplot[only marks,mark=x,mark options={},mark size=1.5000pt,color=black!50!red]
  table[row sep=crcr]{%
39.865	-0.372019251134932\\
40.253	0.827316928424419\\
41.122	-0.130830432656054\\
41.22	0.577868218901744\\
41.691	-0.429838488441512\\
42.024	-0.403406837101361\\
42.78	-0.484353769330574\\
43.06	-0.530609159175838\\
43.867	-0.494265638583131\\
44.125	-0.543824984845914\\
44.968	-0.43644640127655\\
45.216	-0.509133442461965\\
46.162	-0.233253081599139\\
46.288	-0.507481464253206\\
47.269	-0.282812427861923\\
47.432	-0.380279142178729\\
48.122	-0.522349268132041\\
48.457	-0.448010248737866\\
49.235	-0.449662226946626\\
49.543	-0.486005747539333\\
50.295	-0.454618161572904\\
50.594	-0.509133442461965\\
51.39	-0.40836277172764\\
51.66	-0.514089377088244\\
52.418	-0.469485965451739\\
52.698	-0.558692788724748\\
53.509	-0.426534532023994\\
53.784	-0.522349268132041\\
54.628	-0.345587599794781\\
55.196	0.0508871703074838\\
55.858	-0.119266585194738\\
56.236	-0.00858404520785594\\
56.716	-0.443054314111588\\
57.749	0.705070540976221\\
58.11	-0.522349268132041\\
58.597	0.878528252895962\\
58.852	-0.462878052616701\\
59.25	-0.410014749936399\\
59.946	-0.494265638583131\\
60.281	-0.548780919472192\\
61.023	-0.487657725748093\\
61.403	-0.454618161572904\\
62.168	-0.380279142178729\\
62.607	-0.239860994434177\\
63.409	-0.449662226946626\\
63.559	-0.439750357694069\\
64.228	-0.512437398879484\\
64.641	-0.421578597397715\\
65.303	-0.505829486044447\\
65.744	-0.416622662771437\\
66.36	-0.528957180967079\\
66.805	-0.434794423067791\\
67.456	-0.489309703956852\\
67.871	-0.444706292320347\\
68.533	-0.479397834704296\\
68.89	-0.532261137384598\\
70.569	1.12302102779236\\
70.701	-0.443054314111588\\
71.077	-0.448010248737866\\
71.198	1.32456236926101\\
71.789	-0.487657725748093\\
72.147	-0.507481464253206\\
72.867	-0.471137943660498\\
73.299	-0.443054314111588\\
73.971	-0.403406837101361\\
74.352	-0.481049812913055\\
75.407	-0.499221573209409\\
75.496	0.349895226092942\\
76.545	-0.380279142178729\\
76.595	0.38623874668565\\
77.316	-0.238209016225418\\
77.655	-0.312548035619592\\
78.494	-0.129178454447295\\
78.91	-0.0746631735582334\\
79.317	-0.542173006637154\\
80.57	0.896700013192316\\
};
\end{axis}

\pgfplotsset{every axis y label/.append style={rotate=180,yshift=0in,font=\color{blue}}}
\begin{axis}[%
scale=0.2,
width=12.917in,
height=6.647in,
at={(2.167in,0.897in)},
scale only axis,
separate axis lines,
every outer x axis line/.append style={black},
xticklabels={40,45,50,55,60,65,70,75,80},
xtick={40,45,50,55,60,65,70,75,80},
xmin=40,
xmax=80,
xmajorgrids,
every outer y axis line/.append style={blue},
ytick={-2.5,-2.0,-1.5,-1.0,-0.5,0,0.5,1.0,1.5},
every y tick label/.append style={font=\color{blue}},
yticklabels={38,43,48,52,57,62,67,72,76},
ymin=-2.5,
ymax=1.5,
ylabel={\% CPU Utilization},
ymajorgrids,
axis y line*=right,
]
\addplot[solid,mark=o,mark options={},mark size=1.50pt,color=blue,name path=A]
  table[row sep=crcr]{%
39.923	0.15040190523334\\
40.995	1.01077480503545\\
42.086	0.0943837258232653\\
43.1	-0.402946867545325\\
44.113	-0.376769363432766\\
45.129	-0.420811387192308\\
46.143	-0.786246825517428\\
47.16	-0.104792023190334\\
48.173	0.110873244251826\\
49.188	-0.391999338760482\\
50.204	-0.350591859320208\\
51.219	-0.409716911712367\\
52.233	-0.324414355207649\\
53.246	-0.350591859320208\\
54.26	-0.204012530959388\\
55.274	-0.618339233982303\\
56.29	1.04076242702806\\
57.304	-1.21230827176321\\
58.32	1.31686477092576\\
59.334	0.442322011239149\\
60.348	-0.413936381100197\\
61.361	-0.387695899832596\\
62.374	-0.803040733528693\\
63.411	0.469622607949962\\
64.435	-0.336873335713506\\
65.448	-0.235343665592903\\
66.463	-0.240843670466593\\
67.477	-0.193537330837361\\
68.491	-0.31976454192703\\
69.515	-2.13752664887382\\
70.555	1.2103284169793\\
71.583	1.10801153242067\\
72.598	-0.466417343635399\\
73.612	-0.172366510550661\\
74.624	-0.560988038123834\\
75.638	-0.286879970802472\\
76.651	1.16049249495587\\
77.664	-0.791368967460864\\
78.681	0.363201712121078\\
79.7	-0.971767028079368\\
80.759	0.729581807771832\\
};
\path[name path=B] 
        (axis cs:\pgfkeysvalueof{/pgfplots/xmin},\pgfkeysvalueof{/pgfplots/ymax})--
        (axis cs:\pgfkeysvalueof{/pgfplots/xmax},\pgfkeysvalueof{/pgfplots/ymax});
\addplot[gray!30] fill between[of=A and B];
\end{axis}
\end{tikzpicture}%
    \caption{Influence of CPU utilization on service response times.}
    \label{fig:CPUvResp}
    \end{minipage}    
\end{figure}
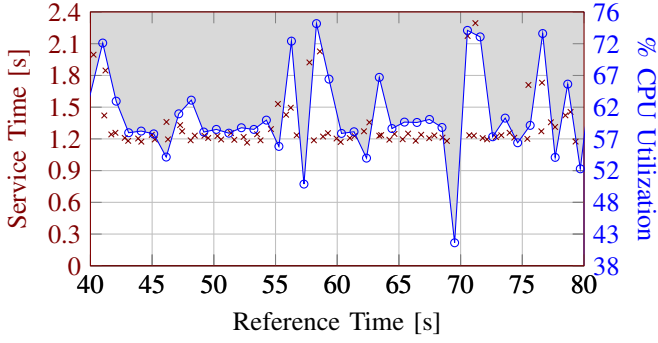

The plot of \eqref{eq:QService} is given in Fig.~\ref{fig:Resp_CPU}. The point at which the gradient of the curve changes from an almost constant value to an exponential rise is chosen as the threshold beyond which service times become unpredictable/unreliable. We choose this by taking the intersection of the tangent to the curve at the point where the gradient is approximately 0.5s per 1\% rise in utilization with the tangent to the curve at the initial point with 0\% utilization. This queuing theory result, though based on the assumption of Poisson arrivals, fits well with our experimental observations made with high entropy in arrival rate. An example of such an observation is depicted in Fig.~\ref{fig:CPUvResp}, where service times are relatively constant around 1.2~s for utilization values lower than 62\%, but vary wildly for higher levels. We define this value as $x_{\mathrm{bnd}}$, the boundary between the predictable and unpredictable regions.

In order to obtain a discretized state space, we partition the utilization values corresponding to predictable response times into regions. To this end, we employ the geometric quantizing function:  
\begin{equation}
    x_{j}= \left \lfloor \left (1 - \left (\dfrac{1}{2}\right )^j \right )x_{\mathrm{tgt}} \right \rfloor, j=0,1,\ldots,n,
    \label{eq:Quant_levels}
\end{equation}
where $x_{\mathrm{tgt}}$ is the ideal utilization level that ensures that resources are highly utilized without compromising system predictability. The expression for $x_j$ is chosen as it results in large steps at the initial stages and smaller steps as the threshold is approached.

In order to ensure that our admission policy keeps the system within the predictable region, we choose the training target, $x_{\mathrm{tgt}}$ as 60\% which is lower than $x_{\mathrm{bnd}}$. By defining $x_n$ as the quantization level closest to the ideal utilization level, we have that $x_n$\textless$x_{\mathrm{tgt}}$\textless$x_{\mathrm{bnd}}$. Operating a VM beyond this region is likely to result in service times that violate SLOs.

By using the geometric quantizer provided in \eqref{eq:Quant_levels}, coarse and fine adjustment is achieved. The former reduces the state space while the latter ensures that the practical utilization limit that still enables admission, $x_\mathrm{lim}$ learned by the system, will be as high as possible.

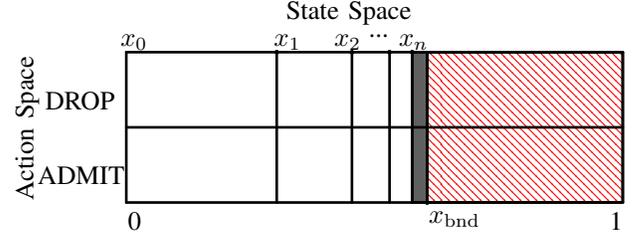
\begin{figure}
    \centering    
    \usetikzlibrary{patterns}
\definecolor{wqwqwq}{rgb}{0.3764705882352941,0.3764705882352941,0.3764705882352941}
\definecolor{ffqqqq}{rgb}{1.,0.,0.}
\begin{tikzpicture}[line cap=round,line join=round,>=triangle 45,x=1.0cm,y=1.0cm,scale=0.2]
\clip(-45.,-7.) rectangle (0.,10.);
\draw[pattern=north west lines, pattern color=ffqqqq] (-15,-5) rectangle (-2,5);
\fill[line width=1.pt,color=wqwqwq,fill=wqwqwq,fill opacity=1.0] (-15.,5.) -- (-16.,5.) -- (-16.,-5.) -- (-15.,-5.) -- cycle;
\draw [line width=1.pt] (-35.,5.)-- (-35.,-5.);
\draw [line width=1.pt] (-25.,5.)-- (-25.,-5.);
\draw [line width=1.pt,dash pattern=on 7pt off 7pt] (-15.,5.)-- (-15.,-5.);
\draw [line width=1.pt] (-20.,5.)-- (-20.,-5.);
\draw [line width=1.pt] (-17.5,5.)-- (-17.5,-5.);
\draw [line width=1.pt] (-2.,5.)-- (-2.,-5.);
\draw [line width=1.pt] (-2.,5.)-- (-35.,5.);
\draw [line width=1.pt] (-35.,-5.)-- (-2.,-5.);
\draw [line width=1.pt] (-15.,5.)-- (-15.,-5.);
\draw [line width=1.pt] (-15.,-5.)-- (-2.,-5.);
\draw [line width=1.pt] (-2.,-5.)-- (-2.,5.);
\draw [line width=1.pt] (-2.,5.)-- (-15.,5.);
\draw [line width=1.pt] (-16.,5.)-- (-16.,-5.);
\draw (-35.5,-5.0) node[anchor=north west] {0};
\draw (-36.0,6.8) node[anchor=north west] {$x_{0}$};
\draw (-25.8,6.8) node[anchor=north west] {$x_{1}$};
\draw (-21.8,6.8) node[anchor=north west] {$x_{2}$};
\draw (-19.5,6.8) node[anchor=north west] {...};
\draw (-17.5,6.8) node[anchor=north west] {$x_{n}$};
\draw (-3.5,-5.0) node[anchor=north west] {1};
\draw (-15.5,-5.0) node[anchor=north west] {$x_{\mathrm{bnd}}$};
\draw [line width=1.pt] (-35.,0.)-- (-2.,0.);
\draw (-41.0,3.0) node[anchor=north west] {DROP};
\draw [line width=1.pt] (-15.,5.)-- (-16.,5.);
\draw [line width=1.pt] (-16.,5.)-- (-16.,-5.);
\draw [line width=1.pt] (-16.,-5.)-- (-15.,-5.);
\draw [line width=1.pt] (-15.,-5.)-- (-15.,5.);
\draw (-41.5,-2.0) node[anchor=north west] {ADMIT};
\draw (-43.0,6.0) node[anchor=north west] {$\rotatebox{90.0}{ \text{Action Space}  }$};
\draw (-25.,9.0) node[anchor=north west] {State Space};
\begin{scriptsize}
\draw [fill=black] (-25.,5.) circle (3.5pt);
\draw [fill=black] (-15.,-5.) circle (3.5pt);
\draw [fill=black] (-20.,5.) circle (3.5pt);
\draw [fill=black] (-17.5,5.) circle (3.5pt);
\draw [fill=black] (-16.,5.) circle (3.5pt);
\end{scriptsize}
\end{tikzpicture}
    \caption{\textit{Q} function Table for training the AC. The gray area represents the ideal operating region at which resources are highly utilized and the service times are within SLOs. The red-shaded area on the right represents the region where VM operation is likely to cause SLO violations.}
    \label{fig:LO_Space}    
\end{figure}

The immediate reward, $R$, for the action taken by the admission controller is the resulting utilization, $x$, discretized to the nearest quantized level boundary~(downwards for a DROP decision or upwards for an ADMIT decision). We do this in order to reduce the variance in the reward values assigned thus accelerating convergence to a given policy. Therefore, with reference to Fig.~\ref{fig:LO_Space}, the reward is calculated as:
\begin{equation}
    R(x)= 
\begin{cases}    
    x_{k},& \text{if DROP}\\    
    x_{k+1},& \text{if ADMIT}, k=0,1,..,n
\end{cases} 
\end{equation}
At the boundary $x_k=x_n,  x_{k+1}=x_{\mathrm{bnd}}$. Beyond the boundary, when $x > x_{\mathrm{bnd}}$ $R(x)$ is defined as,
\begin{equation}
    R(x)= 
\begin{cases}    
    x_{\mathrm{bnd}},& \text{if DROP}\\    
    \frac{1}{2}(x_{\mathrm{bnd}}-1),& \text{if ADMIT},
\end{cases} 
\end{equation}

Note that for cases when $x > x_{\mathrm{bnd}}$, the reward value for ADMIT is negative, implying a penalty for violating the allowable CPU utilization limit.
\subsection{SQLR horizontal scaling agent}
We design and implement a \textit{Q}-Learning scaling agent whose objective is to achieve as low a blocking rate as possible with as few resources as possible. It adds VMs (scale-out) or removes VMs (scale-in) as appropriate given the recent history of utilization experienced by the entire set of active VMs. Therefore the action space for the scaler as a reinforcement learning agent is given by the range of VMs that can be added to or removed from the active set of VMs. The state space consists of three values \begin{enumerate*}[label=(\roman*)]
  \item the current number of VMs, 
  \item the quantized utilization level in the prior epoch and
  \item the quantized utilization level in the current epoch.
\end{enumerate*}

\begin{figure}[t]
    \centering
    \begin{tikzpicture}[line cap=round,line join=round,>=triangle 45,x=1.0cm,y=1.0cm,scale=2.1]
\clip(-11.5,-1.) rectangle (-7.9,2.6);
\draw [line width=1.pt] (-9.8,-0.4)-- (-9.8,0.6);
\draw [line width=1.pt] (-8.8,0.1)-- (-8.8,1.1);
\draw [line width=1.pt] (-3.,0.)-- (-3.,-1.);
\draw [line width=1.pt] (-3.,-1.)-- (-1.8,-0.4);
\draw [line width=1.pt] (-1.8,-0.4)-- (-1.8,0.6);
\draw [line width=1.pt] (-1.8,0.6)-- (-3.,0.);
\draw [line width=1.pt] (-2.6,0.2)-- (-2.6,-0.8);
\draw [line width=1.pt] (-2.2,0.4)-- (-2.2,-0.6);
\draw [line width=1.pt] (-1.2,1.)-- (-1.2,0.);
\draw [line width=1.pt] (-1.2,0.)-- (-0.8,0.2);
\draw [line width=1.pt] (-0.8,0.2)-- (-0.8,1.2);
\draw [line width=1.pt] (-0.8,1.2)-- (-1.2,1.);
\draw [line width=1.pt] (-2.8,0.4)-- (-2.8,0.1);
\draw [line width=1.pt] (-2.8,0.4)-- (-1.6,1.);
\draw [line width=1.pt] (-1.6,1.)-- (-1.6,0.);
\draw [line width=1.pt] (-1.8,-0.1)-- (-1.6,0.);
\draw [line width=1.pt] (-2.4,0.6)-- (-2.4,0.3);
\draw [line width=1.pt] (-2.,0.8)-- (-2.,0.5);
\draw [line width=1.pt] (-1.,1.4)-- (-1.,1.1);
\draw [line width=1.pt] (-1.,1.4)-- (-0.6,1.6);
\draw [line width=1.pt] (-0.6,1.6)-- (-0.6,0.6);
\draw [line width=1.pt] (-0.6,0.6)-- (-0.8,0.5);
\draw [line width=1.pt] (-0.8,2.4)-- (-0.8,1.5);
\draw [line width=1.pt] (-0.6,1.5)-- (-0.4,1.6);
\draw [line width=1.pt] (-0.4,1.6)-- (-0.4,2.6);
\draw [line width=1.pt] (-0.4,2.6)-- (-0.8,2.4);
\draw [line width=1.pt] (-2.6,1.4)-- (-2.6,0.5);
\draw [line width=1.pt] (-1.4,2.)-- (-1.4,1.);
\draw [line width=1.pt] (-2.6,1.4)-- (-1.4,2.);
\draw [line width=1.pt] (-1.6,0.9)-- (-1.4,1.);
\draw [line width=1.pt] (-1.8,1.8)-- (-1.8,0.9);
\draw [line width=1.pt] (-2.2,1.6)-- (-2.2,0.7);
\draw [line width=1.pt] (-1.7,0.85) circle (2.5124689052802225cm);
\draw (-8.776165952910747,1.1556746224066663) node[anchor=north west] {$+ \frac{N}{2}$};
\draw (-10.4,2.35) node[anchor=north west] {$\rotatebox{0.0}{ \text{*State space}}$};
\draw (-10.35,2.07) node[anchor=north west] {$\rotatebox{25.0}{ \text{*$x^{(t+1)}$}}$};
\draw (-10.9,0.3) node[anchor=north west] {$\rotatebox{-45.0}{ \text{Action space}}$};
\draw (-9.8,-0.4) node[anchor=north west] {\{*$K$\}};
\draw [line width=1.pt] (-9.8,0.6)-- (-8.8,1.1);
\draw [line width=1.pt] (-9.8,-0.4)-- (-8.8,0.1);
\draw [line width=1.pt] (-9.8,0.4)-- (-8.8,0.9);
\draw [line width=1.pt] (-9.8,0.2)-- (-8.8,0.7);
\draw [line width=1.pt] (-9.8,0.)-- (-8.8,0.5);
\draw [line width=1.pt] (-9.8,-0.2)-- (-8.8,0.3);
\draw [line width=1.pt] (-9.6,0.7)-- (-9.598533599606084,-0.2992667998030415);
\draw [line width=1.pt] (-9.4,0.8)-- (-9.4,-0.2);
\draw [line width=1.pt] (-9.2,0.9)-- (-9.2,-0.1);
\draw [line width=1.pt] (-9.,1.)-- (-9.,0.);
\draw [line width=1.pt] (-10.1,-0.1)-- (-9.8,0.05);
\draw [line width=1.pt] (-10.1,0.9)-- (-10.1,-0.1);
\draw [line width=1.pt] (-10.1,0.9)-- (-9.1,1.4);
\draw [line width=1.pt] (-9.1,1.4)-- (-9.096671805372653,0.9516640973136741);
\draw [line width=1.pt] (-10.4,0.2)-- (-10.1,0.35);
\draw [line width=1.pt] (-10.4,1.2)-- (-10.4,0.2);
\draw [line width=1.pt] (-10.4,1.2)-- (-9.4,1.7);
\draw [line width=1.pt] (-9.4,1.7)-- (-9.4,1.25);
\draw [line width=1.pt] (-10.1,0.7)-- (-9.098515393833436,1.2000110208846486);
\draw [line width=1.pt] (-10.1,0.5)-- (-9.09703078766687,1.0000220417692973);
\draw [line width=1.pt] (-10.1,0.3)-- (-9.8,0.45);
\draw [line width=1.pt] (-10.1,0.1)-- (-9.8,0.25);
\draw [line width=1.pt] (-9.9,0.)-- (-9.9,1.);
\draw [line width=1.pt] (-9.7,1.1)-- (-9.700652996521631,0.6496735017391847);
\draw [line width=1.pt] (-9.5,1.2)-- (-9.5,0.75);
\draw [line width=1.pt] (-9.3,1.3)-- (-9.3,0.85);
\draw [line width=1.pt] (-10.4,1.)-- (-9.4,1.5);
\draw [line width=1.pt] (-10.4,0.8)-- (-9.4,1.3);
\draw [line width=1.pt] (-10.4,0.6)-- (-10.1,0.75);
\draw [line width=1.pt] (-10.4,0.4)-- (-10.1,0.55);
\draw [line width=1.pt] (-10.2,1.3)-- (-10.2,0.3);
\draw [line width=1.pt] (-10.,1.4)-- (-10.,0.95);
\draw [line width=1.pt] (-9.8,1.5)-- (-9.8,1.05);
\draw [line width=1.pt] (-9.6,1.6)-- (-9.6,1.15);
\draw [line width=1.2pt,dash pattern=on 1pt off 1pt] (-10.56,0.12)-- (-10.44,0.18);
\draw [line width=1.2pt,dash pattern=on 1pt off 1pt] (-9.96,-0.48)-- (-9.84,-0.42);
\draw [->,line width=1.pt] (-10.5,0.15) -- (-10.5,1.15);
\draw [->,line width=1.pt] (-10.5,1.15) -- (-10.5,0.15);
\draw [line width=1.2pt,dash pattern=on 1pt off 1pt] (-10.56,1.12)-- (-10.44,1.18);
\draw (-10.75,0.9) node[anchor=north west] {$\rotatebox{90.0}{ \text{*$x^{(t)}$}}$};

\draw [->,line width=1.pt] (-10.4,1.5) -- (-9.4,2.0);
\draw [->,line width=1.pt] (-9.4,2.0) -- (-10.4,1.5);
\draw (-9.077083031522148,1.4338809781040363) node[anchor=north west] {$0$};
\draw (-9.37800011013355,1.746153418172513) node[anchor=north west] {$- \frac{N}{2}$};
\draw [line width=1.2pt,dash pattern=on 1pt off 1pt] (-10.4,1.6)-- (-10.4,1.4);
\draw [line width=1.2pt,dash pattern=on 1pt off 1pt] (-9.4,1.9)-- (-9.4,2.1);
\draw [->,line width=1.pt] (-10.5,0.15) -- (-9.9,-0.45);
\draw [->,line width=1.pt] (-9.9,-0.45) -- (-10.5,0.15);
\draw [line width=1.pt] (-9.70043501447359,0.8000006307933005) circle (1.6894350114618868cm);
\begin{scriptsize}
\draw [fill=black] (-10.16,0.14) circle (0.5pt);
\draw [fill=black] (-10.2,0.18) circle (0.5pt);
\draw [fill=black] (-10.24,0.22) circle (0.5pt);
\draw [fill=black] (-9.85,-0.15) circle (0.5pt);
\draw [fill=black] (-9.89,-0.11) circle (0.5pt);
\draw [fill=black] (-9.93,-0.07) circle (0.5pt);
\end{scriptsize}
\end{tikzpicture}
    \caption{SQLR action and state space. $K$ is the current number of active VMs, $N$ is the range of VMs that can be added or removed. The state space, whose parameters are prefixed by (*), comprises the number of active VMs and the quantization levels of the average CPU utilization for the set of active VMs.}
    \label{fig:Orch}
\end{figure}
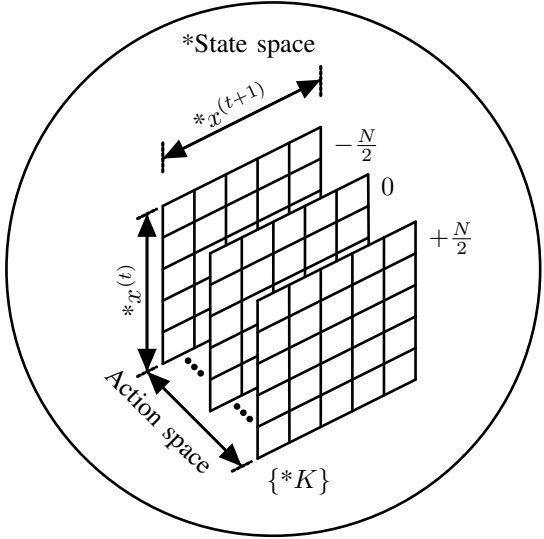

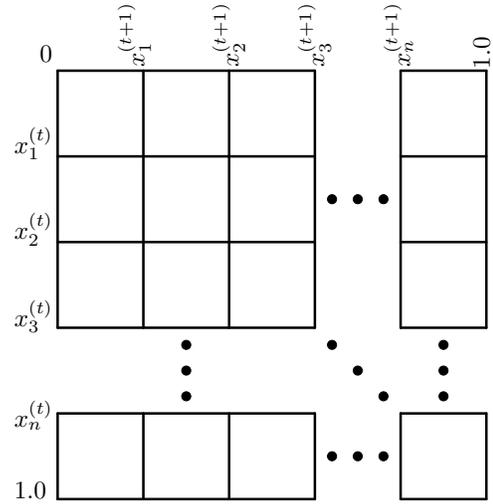
\begin{figure}[t!]
    \centering
    \scalebox{.95}{\begin{tikzpicture}[line cap=round,line join=round,>=triangle 45,x=1.0cm,y=1.0cm,scale=1.2]
\clip(-4.,-2.5) rectangle (2.5,4.);
\draw [line width=1.pt] (-3.,3.)-- (0.,3.);
\draw [line width=1.pt] (-3.,3.)-- (-3.,0.);
\draw [line width=1.pt] (-3.,-1.)-- (-3.,-2.);
\draw [line width=1.pt] (1.,3.)-- (2.,3.);
\draw [line width=1.pt] (1.,-2.)-- (2.,-2.);
\draw [line width=1.pt] (2.,-1.)-- (2.,-2.);
\draw [line width=1.pt] (1.,-1.)-- (2.,-1.);
\draw [line width=1.pt] (1.,-1.)-- (1.,-2.);
\draw [line width=1.pt] (0.,-1.)-- (0.,-2.);
\draw [line width=1.pt] (-3.,-2.)-- (0.,-2.);
\draw [line width=1.pt] (-3.,-1.)-- (0.,-1.);
\draw [line width=1.pt] (-3.,0.)-- (0.,0.);
\draw [line width=1.pt] (0.,3.)-- (0.,0.);
\draw [line width=1.pt] (1.,3.)-- (1.,0.);
\draw [line width=1.pt] (2.,3.)-- (2.,0.);
\draw [line width=1.pt] (1.,0.)-- (2.,0.);
\draw [line width=1.pt] (-3.,2.)-- (0.,2.);
\draw [line width=1.pt] (-3.,1.)-- (0.,1.);
\draw [line width=1.pt] (-2.,-1.)-- (-2.,-2.);
\draw [line width=1.pt] (-1.,-1.)-- (-1.,-2.);
\draw [line width=1.pt] (1.,2.)-- (2.,2.);
\draw [line width=1.pt] (1.,1.)-- (2.,1.);
\draw [line width=1.pt] (-2.,3.)-- (-2.,0.);
\draw [line width=1.pt] (-1.,3.)-- (-1.,0.);
\draw (-3.31,3.4) node[anchor=north west] {$0$};
\draw (-3.61,2.45) node[anchor=north west] {$x^{(t)}_{1}$};
\draw (-3.61,1.45) node[anchor=north west] {$x^{(t)}_{2}$};
\draw (-3.61,0.43) node[anchor=north west] {$x^{(t)}_{3}$};
\draw (-3.61,-0.75) node[anchor=north west] {$x^{(t)}_{n}$};
\draw (-3.61,-1.70) node[anchor=north west] {$1.0$};
\draw (-2.40,3.90) node[anchor=north west] {$\rotatebox{90.0}{{$x^{(t+1)}_{1}$}}$};
\draw (-1.3,3.90) node[anchor=north west] {$\rotatebox{90.0}{{$x^{(t+1)}_{2}$}}$};
\draw (-0.3,3.90) node[anchor=north west] {$\rotatebox{90.0}{{$x^{(t+1)}_{3}$}}$};
\draw (0.7,3.90) node[anchor=north west] {$\rotatebox{90.0}{{$x^{(t+1)}_{n}$}}$};
\draw (1.75,3.50) node[anchor=north west] {$\rotatebox{90.0}{{$1.0$}}$};
\begin{scriptsize}
\draw [fill=black] (-1.5,-0.2) circle (1.5pt);
\draw [fill=black] (-1.5,-0.8) circle (1.5pt);
\draw [fill=black] (-1.5,-0.5) circle (1.5pt);
\draw [fill=black] (0.2,1.5) circle (1.5pt);
\draw [fill=black] (0.5,1.5) circle (1.5pt);
\draw [fill=black] (0.8,1.5) circle (1.5pt);
\draw [fill=black] (0.8,-0.8) circle (1.5pt);
\draw [fill=black] (0.2,-0.2) circle (1.5pt);
\draw [fill=black] (0.5,-0.5) circle (1.5pt);
\draw [fill=black] (1.5,-0.2) circle (1.5pt);
\draw [fill=black] (1.5,-0.5) circle (1.5pt);
\draw [fill=black] (1.5,-0.8) circle (1.5pt);
\draw [fill=black] (0.2,-1.5) circle (1.5pt);
\draw [fill=black] (0.5,-1.5) circle (1.5pt);
\draw [fill=black] (0.8,-1.5) circle (1.5pt);
\end{scriptsize}
\end{tikzpicture}}
    \caption{State space detail for a \textquotedblleft card\textquotedblright\ in the action space. Each cell\textquoteright s index pair is given by the quantized level of average system-wide resource utilization in successive episodes.}
    \label{fig:State_space}
\end{figure}

The state and action spaces for our horizontal scaling agent are as shown in Fig.~\ref{fig:Orch}. Each permissible action is represented as a \textquotedblleft card\textquotedblright\ indicating the number of VMs that need to be added or removed in case of taking the action associated with the card. Moreover, each card consists of a grid whose rows and columns are indexed with load levels (to be interpreted as the immediate past and current quantized load, respectively), as highlighted in Fig.~\ref{fig:State_space}. The cells contain the cumulative reward obtained by a given state-action pair. We use a uniform quantizer for the state space since it provides a more granular view of the level of system-wide resource utilization than the geometric quantizer used for the admission control agent. However, because the context of a given utilization level is provided by how many active nodes are considered, greater nuance is required to distinguish the state space, at the lower levels. To do so, we choose smaller steps (2\%) in the region between 0-20\% of utilization and 5\% in the region between 20\% and $x_\mathrm{lim}$. The region above $x_\mathrm{lim}$ is chosen as one large level, given that at this region of utilization a coarse scaling decision is most likely, it does not require a fine resolution in the state space representation.

\subsubsection{The Action Space}
The scaling reward function~($R_{\mathrm{sqlr}}$) consists of two components: $R_{\mathrm{blk}}$ computed using the blocking probability $P$ and $R_{\mathrm{res}}$ computed from the resource cost (dependent on the number of active VMs, $K$). 
\begin{equation}
\begin{split}
R_{\mathrm{sqlr}}&= R_{\mathrm{blk}}+R_{\mathrm{res}}\\
R_{\mathrm{blk}}&= 
\begin{cases}
    R_{\mathrm{min}},& \text{if } P\leqslant P_{\mathrm{blk}}\\
    \theta\left(P_{\mathrm{blk}}-P\right),& \text{if } P > P_{\mathrm{blk}},
\end{cases}\\
R_{\mathrm{res}}&=\beta(1-K)
\end{split}
\label{eq:Reward_params}
\end{equation}
where $P_{\mathrm{blk}}$ is the ideal blocking probability threshold and $R_{\mathrm{min}}$ is a small positive reward given to the agent as an incentive for keeping the system within the allowable service outage limits. The training parameters $\theta$ and $\beta$ act as modifiers, so that blocking probability violations receive a different penalty than that brought about by usage of extra resources. The reason is that the financial penalties incurred by a CSP for violating SLOs may be different from cost savings achieved by reducing resource usage.

\begin{figure}[t!]
    \centering
    \includegraphics{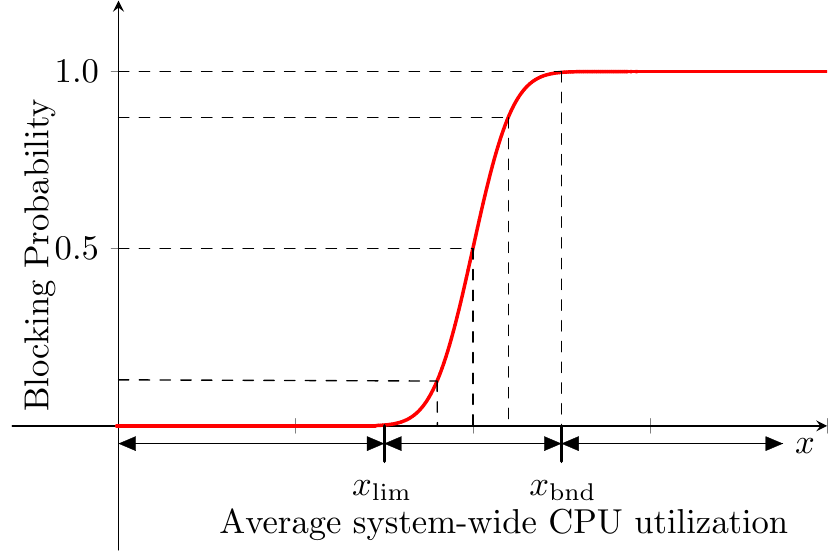}
    \caption{Modified error function to estimate the blocking probability component of the initial $Q$ values of card \textquotedblleft 0\textquotedblright\ (Fig.~\ref{fig:Orch}) diagonals.}
    \label{fig:Pb_Init}
\end{figure}

\subsubsection{The State Space}
\begin{figure}[t]
    \centering
    \input{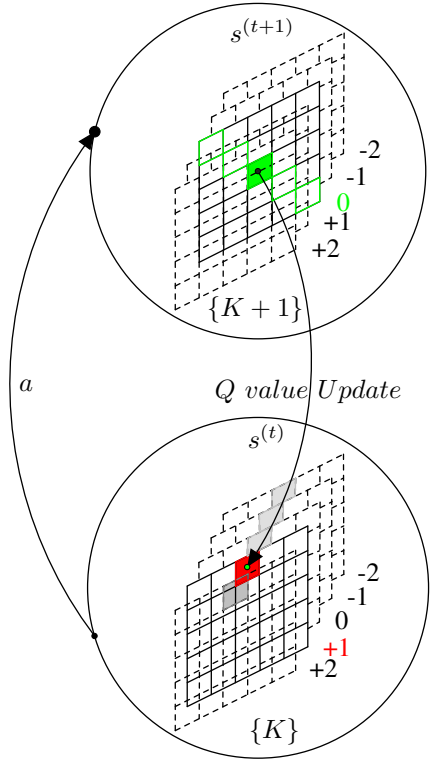}
    \caption{SQLR horizontal scaling mechanism. The grey-shaded cells are those whose \textit{Q}-values are compared to determine the action to take according to \eqref{eq:selectioner}. A scale OUT of \textquotedblleft $+1$\textquotedblright\ is chosen as the action. After scaling action $a$, the \textit{Q}-value in the red-shaded cell receives the update as specified by \eqref{eq:modified_Q}. One component of the update is the \textit{Q}-value contained in the green-shaded cell of card \textquotedblleft 0\textquotedblright\ in bubble \textquotedblleft $\{K+1\}$\textquotedblright.}%
    \label{fig:Horizontal_Scaling}%
\end{figure}

As stated earlier, each card in the bubble shown in Fig.~\ref{fig:Orch} consists of a grid whose cell indices correspond to the average level of utilization of the active VMs over the previous epoch and the current epoch (cf. Fig.~\ref{fig:State_space}). The diagonal elements for the cases where the number of VMs remain unchanged~(card \textquotedblleft 0\textquotedblright\ in Fig.~\ref{fig:Orch}) are initialized as non-zero values. We do this on the basis of the cost associated with the number of active nodes and estimated values of blocking probability, $P_{0}$, derived from the error function as shown in Fig.~\ref{fig:Pb_Init}:
\begin{align}
    P_{0}(x)&= 
\begin{cases}
    0,& x<x_\mathrm{lim}\\
    1,& x>x_{\mathrm{bnd}}\\
   \dfrac{1}{2}\left [1+\erf\left (\eta(x)\dfrac{\rm{e}}{\sqrt{2}} \right ) \right ],&  \mathrm{otherwise}
\end{cases}
\label{eq:Init_P}
\end{align}\\
where (cf. Fig.~\ref{fig:Pb_Init})
\begin{equation}
\eta(x) =\dfrac{x-x_\mathrm{lim}}{x_{\mathrm{bnd}}-x_\mathrm{lim}},
\end{equation}
and we recall that $x_\mathrm{lim}$ is learned adaptively by the admission control agent.
These diagonal elements serve as the reference action values,~$Q(S^{(t+1)}, a)$, for the updates  in \eqref{eq:modified_Q} after horizontal scaling. This process is depicted in Fig.~\ref{fig:Horizontal_Scaling} for a scale-out action. To describe this, we consider starting after an episodic wait period, the previous action having taken place at instant $t-1$. The first cell index is the quantized level of the average utilization in the interval $[t-2,t-1)$. At instance $t$ our scaler obtains the quantized level of the average utilization in the interval $[t-1,t)$. This serves as the second cell index to be considered in selecting the action. The current number of active VMs, $K$, is also evaluated. With this triplet of values, the current state is established. Then, by leveraging the \textit{Q}-value entries of the cells with this reference set of indices in every card of the action space in the bubble defined by $K$~VMs, a scaling action is chosen based on \eqref{eq:selectioner}.

For later reference we term this cell, in the chosen action card, as R-Cell~(marked red in Fig.~\ref{fig:Horizontal_Scaling}). After waiting a short period for the VMs to start-up or shut-down, a further predefined wait period between episodes is observed to allow the effect of the change to be manifest. We are now at instant $t+1$. The immediate reward value is then calculated by taking account of the blocking probability observed between time instants $t$ and $t+1$ and of the number of active VMs at instant $t+1$ as described in \eqref{eq:Reward_params}. We also take into account the accumulated reward stored in card \textquotedblleft 0\textquotedblright\ at the diagonal cell whose two indices are both given by the quantized average utilization values over the interval $[t,t+1)$~(marked green in Fig.~\ref{fig:Horizontal_Scaling}). These two values are used to update the value in R-Cell as prescribed in \eqref{eq:modified_Q}. Recall that the diagonal cells in card \textquotedblleft 0\textquotedblright\ are initialized assuming that the average utilization in successive epochs is stable and their \textit{Q}-values can be estimated using \eqref{eq:Reward_params} and \eqref{eq:Init_P}.
\section{Experimental Setup}
\label{experimentum}

\begin{figure}[t]
  \centering
\includegraphics[scale=0.53]{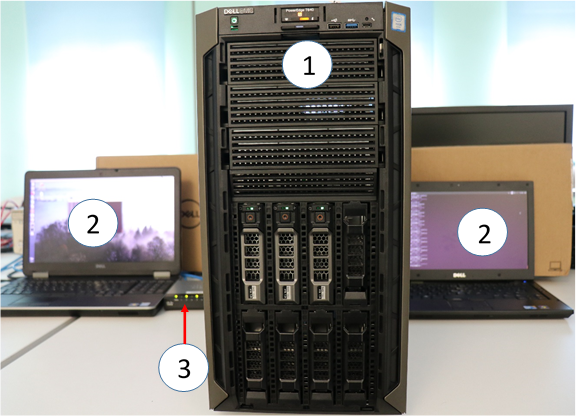}
\caption{Testbed setup. (1)~Dell T640 server (2)~Client PCs (3)~Gbps switch.}
  \label{fig:LiveLAN}
\end{figure}

In order to evaluate the effectiveness of our scheme we run experiments on a test-bed that mirrors the operations of a CSP. We setup the test-bed as shown in Fig.~\ref{fig:LiveLAN}. The processor architecture of our Dell T640 server consists of two processor sockets with Non Uniform Memory Allocation~(NUMA), 10 hyper-threaded CPU cores per socket for a total of 40 logical cores with a variable clock rate. The server memory is 128~GB. 

The server runs Ubuntu 18.04 LTS as its operating system and acts as a host for virtual machines. The Client PCs run on Ubuntu 16.04.3 LTS. We use KVM as the hypervisor and manage the virtual machines using libvirt~\cite{LibVirt}. Each instance of a virtual machine is configured with 4 virtual CPUs and 4~GB of memory. The client PCs and the server are connected via a Cisco switch to form a Gigabit/s local area network. The PCs function as ASPs running bash scripts that generate requests to the server with varying rates as depicted in Figs.~\ref{fig:Training_Load} and \ref{fig:Offered_Load}.

The server launches VMs to handle incoming requests according to one of the following schemes: static provisioning, extended Kalman filtering based prediction, and our proposed SQLR. All schemes including the admission control and load balancer VNFs are implemented in Python and are run within the host.

As mentioned in Section \ref{relatio}, the state-of-the-art scheme in \cite{Gandhi} leverages a queuing system model enhanced with an Extended Kalman Filter~(EKF). It makes near time predictions of response times based on measurements of arrival rates and system utilization. Using a queue model refined by a tuned EKF with the maximum allowable response time~(from an SLA) as input, it then calculates the number of nodes needed and scales appropriately to get to this number.

We make some slight modifications to the EKF algorithm to make it more robust. We increase the interval between the predict and update phases from 10s to 90s. This provides sufficient time for starting up a VM and letting it handle traffic. Additionally, instead of the instantaneous measured system utilization and response times, we provide their average over the predict and update intervals of the filter as input to the EKF. This prevents the scaler from over/under estimating input parameters, and thus yields a fairer comparison to SQLR. Further, we dispense with the network delay in the system model as the response times are taken directly on the server. We consider a single-tiered application, and one class of requests. This also has the effect of simplifying the process and measurement of the noise covariance matrices to $\mathbb{R}^{2\times 2}$ (as only two parameters are taken into account in each case), thereby enhancing the tuning of the EKF.

For SQLR, we limit the maximum number of virtual machines that can be added or removed to 2 for our experiments. This reduces the state space and the number of episodes required for convergence,. In addition, we employ a damping mechanism which (for the fully converged states) requires two consecutive scale-in decisions before withdrawing resources. The latter ameliorates premature removal of resources while allowing expedited provisioning to handle sudden increases in demand.

We choose as our cloud application the double 256-bit hashing algorithm used for proof-of-work computation in bit-coin mining~\cite{Bitcoin}. It is a suitable stand-in for resource-hungry, computationally challenging tasks that are commonly deferred to the cloud such as encryption~\cite{EncryptionRef} and transcoding~\cite{TranscodingRef}, \cite{Ayimba:2019}. Each iteration of this computation involves incrementing a counter variable (nonce) which is then hashed together with a given hash code and merkle root. The output is then hashed again. The hashing mechanism employed is the 256-bit Secure Hash Algorithm~(SHA-256). We will use the word \textit{job} to refer to one proof-of-work iteration from hereon.

We limit the number of iterations generated by a request to specific target values so as to mimic the varying degrees of complexity of typical cloud application requests. The number of iterations generated by a request comes from the discrete set \{300k, 400k,\ldots,1200k\}. 

\begin{figure}[t]
    \centering
    \input{TraffiQ2_profile.tex}    
    \caption{Pre-training workload profile. The red line is the moving average of requests with a window of 30 samples.}
    \label{fig:Training_Load}
    \vspace{1.5em}%
    \input{TraffiQ_profile.tex}    
    \caption{Test workload profile. The red line is the moving average of requests with a window of 30 samples.}
    \label{fig:Offered_Load}
\end{figure}
As part of the training for the SQLR elastic provisioning system, we combine several traffic profiles with various workload averages, resulting in the composite shown in Fig.~\ref{fig:Training_Load}.

For the test traffic, we again use a combination of several profiles with different workload averages to obtain the composite shown in Fig.~\ref{fig:Offered_Load}. To achieve this we configure requests to be sent with inter-arrival times of the discrete distribution $\omega \sim \mathcal {U}\{0,\omega{\mathrm{max}}\}$ for each hour slot. For example for the busy-hour slot $\omega_{\mathrm{max}} = 5\mathrm{s}$ and the for low traffic period  $\omega_{\mathrm{max}} = 9\mathrm{s}$. This results in high entropy (given the uniform distribution of inter-arrival times) and presents greater stochastic behaviour which is more challenging than those encountered in common reference profiles as those used in~\cite{Gandhi}. Moreover, it exhibits patterns encountered in real workloads, with rapid variations over short intervals, but with veritable trends over longer observation windows. It also includes sudden bursts and drops such as those observed at the start of hours 7, 10, 14 and 18. We run multiple instances of the admission control and load balancer functions, which increases the volume of requests received at the server side. Simply increasing the rate of arrivals results in port binding errors and low numbers of requests getting through.
\section{Results}
\label{resultet}
In this section, we show the effectiveness of policy convergence leveraging the results from the training of the admission control function. We then examine the results with respect to two SLOs: service availability (as measured via blocking rates) and response times.
\subsection{Admission control}
We briefly discuss our admission control function. We recall that this function learns the appropriate utilization limit, $x_\mathrm{lim}$, to be applied in its policy in order to ensure that response times are reliably bounded.
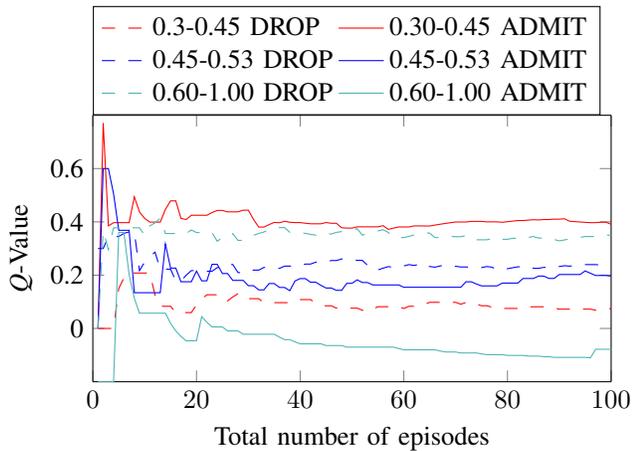
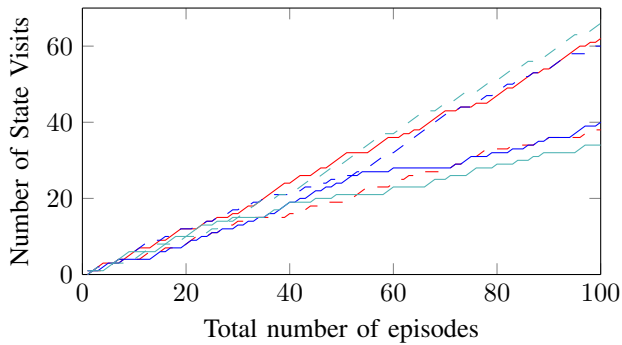
\begin{figure}%
    \centering
    \subfloat[Reward accumulated with experience.]{{\usetikzlibrary{decorations}
\usetikzlibrary{decorations.markings}
\makeatletter
\tikzset{
  nomorepostactions/.code={\let\tikz@postactions=\pgfutil@empty},
  mymark/.style 2 args={decoration={markings,
    mark= between positions 0 and 1 step (1/11)*\pgfdecoratedpathlength with{%
        \tikzset{#2,every mark}\tikz@options
        \pgfuseplotmark{#1}%
      },  
    },
    postaction={decorate},
    /pgfplots/legend image post style={
            mark=#1,#2,every path/.append style={nomorepostactions}
    },
  },
}
\makeatother
\definecolor{mycolor1}{rgb}{0.30000,0.70000,0.70000}%
\begin{tikzpicture}
\begin{axis}[%
width=12.917in,
height=6.647in,
scale=0.21,
at={(1.85in,0.746in)},
scale only axis,
separate axis lines,
every outer x axis line/.append style={black},
every x tick label/.append style={font=\color{black}},
xmin=0,
xmax=100,
xlabel={Total number of episodes},
every outer y axis line/.append style={black},
every y tick label/.append
style={font=\color{black}},
yticklabel style={
    /pgf/number format/fixed,
    /pgf/number format/precision=3
},
scaled y ticks=false,
ymin=-0.2,
ymax=0.8,
ytick={0,0.2,0.4,0.6},
yticklabels={0,0.2,0.4,0.6},
ylabel={\textit{Q}-Value},
axis background/.style={fill=white},
legend style={at={(0.0,1.0)},anchor=south west,draw=none, legend columns=2,legend cell align=left,align=left,draw=black}
]
\addplot [color=red,dash pattern=on 5pt off 5pt]
  table[row sep=crcr]{%
1	0\\
2	0\\
3	0\\
4	0\\
5	0.15\\
6	0.19\\
7	0.208\\
8	0.208\\
9	0.208\\
10	0.208\\
11	0.208\\
12	0.1248\\
13	0.0832\\
14	0.0832\\
15	0.0832\\
16	0.0594\\
17	0.0594\\
18	0.0594\\
19	0.0594\\
20	0.088\\
21	0.1096\\
22	0.1264\\
23	0.1264\\
24	0.1264\\
25	0.1034\\
26	0.1034\\
27	0.1181\\
28	0.1303\\
29	0.1303\\
30	0.1117\\
31	0.1117\\
32	0.1117\\
33	0.0968\\
34	0.0968\\
35	0.0968\\
36	0.0968\\
37	0.0968\\
38	0.0968\\
39	0.0968\\
40	0.1083\\
41	0.1083\\
42	0.1083\\
43	0.0956\\
44	0.085\\
45	0.085\\
46	0.085\\
47	0.0761\\
48	0.0761\\
49	0.0761\\
50	0.0761\\
51	0.0761\\
52	0.0685\\
53	0.062\\
54	0.0723\\
55	0.0816\\
56	0.0816\\
57	0.0816\\
58	0.0816\\
59	0.0816\\
60	0.0748\\
61	0.0832\\
62	0.0832\\
63	0.0909\\
64	0.0909\\
65	0.098\\
66	0.098\\
67	0.098\\
68	0.098\\
69	0.098\\
70	0.098\\
71	0.091\\
72	0.0976\\
73	0.0976\\
74	0.0911\\
75	0.0852\\
76	0.0852\\
77	0.0799\\
78	0.0751\\
79	0.0751\\
80	0.0751\\
81	0.0751\\
82	0.0751\\
83	0.0813\\
84	0.0813\\
85	0.0813\\
86	0.0813\\
87	0.0813\\
88	0.0767\\
89	0.0767\\
90	0.0724\\
91	0.0724\\
92	0.0724\\
93	0.0724\\
94	0.0724\\
95	0.0724\\
96	0.0724\\
97	0.0685\\
98	0.0685\\
99	0.0742\\
100	0.0742\\
};
\addlegendentry{0.3-0.45 DROP};
\addplot [color=red,width=1.0,solid]
  table[row sep=crcr]{%
1	0\\
2	0.77\\
3	0.385\\
4	0.397\\
5	0.397\\
6	0.397\\
7	0.397\\
8	0.4937\\
9	0.437\\
10	0.4129\\
11	0.3991\\
12	0.3991\\
13	0.3991\\
14	0.4469\\
15	0.4788\\
16	0.4788\\
17	0.413\\
18	0.4089\\
19	0.4247\\
20	0.4247\\
21	0.4247\\
22	0.4247\\
23	0.4349\\
24	0.4429\\
25	0.4429\\
26	0.4375\\
27	0.4375\\
28	0.4375\\
29	0.4438\\
30	0.4438\\
31	0.4092\\
32	0.3804\\
33	0.3804\\
34	0.3898\\
35	0.3979\\
36	0.3966\\
37	0.4022\\
38	0.3984\\
39	0.3972\\
40	0.3972\\
41	0.3948\\
42	0.3927\\
43	0.3927\\
44	0.3927\\
45	0.3975\\
46	0.3947\\
47	0.3947\\
48	0.3778\\
49	0.3766\\
50	0.3765\\
51	0.3813\\
52	0.3813\\
53	0.3813\\
54	0.3813\\
55	0.3813\\
56	0.3857\\
57	0.3718\\
58	0.3765\\
59	0.3808\\
60	0.3808\\
61	0.3808\\
62	0.3848\\
63	0.3848\\
64	0.3845\\
65	0.3845\\
66	0.3879\\
67	0.3911\\
68	0.394\\
69	0.3968\\
70	0.3994\\
71	0.3994\\
72	0.3994\\
73	0.4018\\
74	0.4018\\
75	0.4018\\
76	0.4\\
77	0.4\\
78	0.4\\
79	0.3983\\
80	0.4006\\
81	0.4028\\
82	0.4049\\
83	0.4049\\
84	0.4065\\
85	0.408\\
86	0.4072\\
87	0.4087\\
88	0.4087\\
89	0.4101\\
90	0.4101\\
91	0.4115\\
92	0.4022\\
93	0.4008\\
94	0.3995\\
95	0.3982\\
96	0.397\\
97	0.397\\
98	0.3987\\
99	0.3987\\
100	0.3907\\
};
\addlegendentry{0.30-0.45 ADMIT};
\addplot [color=blue,dash pattern=on 5pt off 5pt]
  table[row sep=crcr]{%
1	0.3\\
2	0.3\\
3	0.345\\
4	0.345\\
5	0.345\\
6	0.357\\
7	0.3624\\
8	0.3624\\
9	0.2174\\
10	0.2489\\
11	0.2705\\
12	0.2862\\
13	0.2226\\
14	0.2226\\
15	0.2226\\
16	0.1781\\
17	0.1781\\
18	0.1996\\
19	0.2171\\
20	0.2171\\
21	0.2171\\
22	0.2146\\
23	0.2146\\
24	0.2283\\
25	0.2283\\
26	0.24\\
27	0.24\\
28	0.21\\
29	0.2078\\
30	0.2078\\
31	0.2189\\
32	0.2189\\
33	0.2288\\
34	0.2288\\
35	0.2254\\
36	0.2254\\
37	0.2339\\
38	0.2339\\
39	0.2339\\
40	0.2339\\
41	0.2416\\
42	0.2416\\
43	0.2486\\
44	0.2486\\
45	0.2486\\
46	0.2549\\
47	0.2549\\
48	0.2607\\
49	0.2607\\
50	0.2555\\
51	0.2555\\
52	0.2555\\
53	0.2366\\
54	0.2366\\
55	0.2197\\
56	0.2261\\
57	0.2321\\
58	0.229\\
59	0.2345\\
60	0.2345\\
61	0.2396\\
62	0.2363\\
63	0.2411\\
64	0.2277\\
65	0.2325\\
66	0.2299\\
67	0.2344\\
68	0.2386\\
69	0.2358\\
70	0.2332\\
71	0.2307\\
72	0.2307\\
73	0.2284\\
74	0.2284\\
75	0.2284\\
76	0.2323\\
77	0.23\\
78	0.2202\\
79	0.2202\\
80	0.2241\\
81	0.2278\\
82	0.2313\\
83	0.2313\\
84	0.2347\\
85	0.2257\\
86	0.2257\\
87	0.2291\\
88	0.2291\\
89	0.2323\\
90	0.2323\\
91	0.2354\\
92	0.2337\\
93	0.2367\\
94	0.2396\\
95	0.2396\\
96	0.2396\\
97	0.2396\\
98	0.2379\\
99	0.2406\\
100	0.2406\\
};
\addlegendentry{0.45-0.53 DROP};
\addplot [color=blue,width=1.0,solid]
  table[row sep=crcr]{%
1	0\\
2	0.6\\
3	0.6\\
4	0.505\\
5	0.3683\\
6	0.3683\\
7	0.3683\\
8	0.1341\\
9	0.1341\\
10	0.1341\\
11	0.1341\\
12	0.1341\\
13	0.1341\\
14	0.3166\\
15	0.2266\\
16	0.2266\\
17	0.1752\\
18	0.1752\\
19	0.1752\\
20	0.2152\\
21	0.1787\\
22	0.1787\\
23	0.2406\\
24	0.2406\\
25	0.2061\\
26	0.2061\\
27	0.1803\\
28	0.1803\\
29	0.1803\\
30	0.1604\\
31	0.1604\\
32	0.1448\\
33	0.1448\\
34	0.1685\\
35	0.1685\\
36	0.1538\\
37	0.1538\\
38	0.1417\\
39	0.1781\\
40	0.1871\\
41	0.1871\\
42	0.1735\\
43	0.1735\\
44	0.1618\\
45	0.1517\\
46	0.1517\\
47	0.143\\
48	0.143\\
49	0.1693\\
50	0.1693\\
51	0.1824\\
52	0.1723\\
53	0.1723\\
54	0.1633\\
55	0.1633\\
56	0.1633\\
57	0.1633\\
58	0.1633\\
59	0.1633\\
60	0.1545\\
61	0.1545\\
62	0.1545\\
63	0.1545\\
64	0.1545\\
65	0.1545\\
66	0.1545\\
67	0.1545\\
68	0.1545\\
69	0.1545\\
70	0.1545\\
71	0.1545\\
72	0.1751\\
73	0.1751\\
74	0.167\\
75	0.1597\\
76	0.1597\\
77	0.1597\\
78	0.1597\\
79	0.1703\\
80	0.1703\\
81	0.1703\\
82	0.1703\\
83	0.1875\\
84	0.1875\\
85	0.1875\\
86	0.1965\\
87	0.1965\\
88	0.1877\\
89	0.1877\\
90	0.2024\\
91	0.2024\\
92	0.2024\\
93	0.2024\\
94	0.2024\\
95	0.2159\\
96	0.2071\\
97	0.1987\\
98	0.1987\\
99	0.1987\\
100	0.1963\\
};
\addlegendentry{0.45-0.53 ADMIT};
\addplot [color=mycolor1,dash pattern=on 5pt off 5pt]
  table[row sep=crcr]{%
1	0\\
2	0.3666\\
3	0.2962\\
4	0.3777\\
5	0.3777\\
6	0.3777\\
7	0.3777\\
8	0.3777\\
9	0.3777\\
10	0.3487\\
11	0.385\\
12	0.3962\\
13	0.414\\
14	0.3562\\
15	0.3562\\
16	0.3562\\
17	0.3562\\
18	0.3562\\
19	0.3696\\
20	0.3598\\
21	0.3598\\
22	0.3598\\
23	0.3598\\
24	0.3271\\
25	0.3411\\
26	0.3411\\
27	0.3558\\
28	0.3305\\
29	0.3305\\
30	0.3414\\
31	0.3496\\
32	0.3563\\
33	0.3619\\
34	0.3706\\
35	0.3784\\
36	0.3784\\
37	0.3784\\
38	0.3607\\
39	0.3607\\
40	0.3607\\
41	0.3567\\
42	0.3532\\
43	0.35\\
44	0.3572\\
45	0.3572\\
46	0.3542\\
47	0.3607\\
48	0.3634\\
49	0.3634\\
50	0.3691\\
51	0.3743\\
52	0.3502\\
53	0.3558\\
54	0.361\\
55	0.3659\\
56	0.3705\\
57	0.3604\\
58	0.3409\\
59	0.3409\\
60	0.3409\\
61	0.3459\\
62	0.3506\\
63	0.3551\\
64	0.3593\\
65	0.3576\\
66	0.341\\
67	0.341\\
68	0.341\\
69	0.3453\\
70	0.3443\\
71	0.3443\\
72	0.3484\\
73	0.3472\\
74	0.3327\\
75	0.3327\\
76	0.3327\\
77	0.3368\\
78	0.3407\\
79	0.3346\\
80	0.3346\\
81	0.3384\\
82	0.3378\\
83	0.3414\\
84	0.3414\\
85	0.3449\\
86	0.3392\\
87	0.3392\\
88	0.3338\\
89	0.3338\\
90	0.3287\\
91	0.3322\\
92	0.3356\\
93	0.3388\\
94	0.3419\\
95	0.3449\\
96	0.3449\\
97	0.3449\\
98	0.3478\\
99	0.3506\\
100	0.3495\\
};
\addlegendentry{0.60-1.00 DROP};
\addplot [color=mycolor1,width=1.0,solid]
  table[row sep=crcr]{%
1	-0.2\\
2	-0.2\\
3	-0.2\\
4	-0.2\\
5	0.361\\
6	0.3594\\
7	0.2016\\
8	0.1132\\
9	0.0572\\
10	0.0572\\
11	0.0572\\
12	0.0572\\
13	0.0572\\
14	0.0572\\
15	0.0188\\
16	-0.009\\
17	-0.03\\
18	-0.0464\\
19	-0.0464\\
20	-0.0464\\
21	0.0443\\
22	0.0232\\
23	0.0057\\
24	0.0057\\
25	0.0057\\
26	-0.0091\\
27	-0.0091\\
28	-0.0091\\
29	-0.0217\\
30	-0.0217\\
31	-0.0217\\
32	-0.0217\\
33	-0.0217\\
34	-0.0217\\
35	-0.0217\\
36	-0.0326\\
37	-0.0421\\
38	-0.0421\\
39	-0.0504\\
40	-0.0577\\
41	-0.0577\\
42	-0.0577\\
43	-0.0577\\
44	-0.0577\\
45	-0.0642\\
46	-0.0642\\
47	-0.0642\\
48	-0.0642\\
49	-0.0701\\
50	-0.0701\\
51	-0.0701\\
52	-0.0701\\
53	-0.0701\\
54	-0.0701\\
55	-0.0701\\
56	-0.0701\\
57	-0.0701\\
58	-0.0701\\
59	-0.0754\\
60	-0.0802\\
61	-0.0802\\
62	-0.0802\\
63	-0.0802\\
64	-0.0802\\
65	-0.0802\\
66	-0.0802\\
67	-0.0845\\
68	-0.0884\\
69	-0.0884\\
70	-0.0884\\
71	-0.092\\
72	-0.092\\
73	-0.092\\
74	-0.092\\
75	-0.0953\\
76	-0.0984\\
77	-0.0984\\
78	-0.0984\\
79	-0.0984\\
80	-0.1012\\
81	-0.1012\\
82	-0.1012\\
83	-0.1012\\
84	-0.1038\\
85	-0.1038\\
86	-0.1038\\
87	-0.1062\\
88	-0.1062\\
89	-0.1085\\
90	-0.1085\\
91	-0.1085\\
92	-0.1085\\
93	-0.1085\\
94	-0.1085\\
95	-0.1085\\
96	-0.1106\\
97	-0.0781\\
98	-0.0781\\
99	-0.0781\\
100	-0.0781\\
};
\addlegendentry{0.60-1.00 ADMIT};

\end{axis}
\end{tikzpicture}%
 \label{fig:AC_Learn_1} }}%
    \qquad
    \subfloat[Frequency of DROP and ADMIT decisions. ]{{\usetikzlibrary{decorations}
\usetikzlibrary{decorations.markings}
\makeatletter
\tikzset{
  nomorepostactions/.code={\let\tikz@postactions=\pgfutil@empty},
  mymark/.style 2 args={decoration={markings,
    mark= between positions 0 and 1 step (1/11)*\pgfdecoratedpathlength with{%
        \tikzset{#2,every mark}\tikz@options
        \pgfuseplotmark{#1}%
      },  
    },
    postaction={decorate},
    /pgfplots/legend image post style={
            mark=#1,#2,every path/.append style={nomorepostactions}
    },
  },
}
\makeatother
\definecolor{mycolor1}{rgb}{0.30000,0.70000,0.70000}%
\begin{tikzpicture}

\begin{axis}[%
width=12.917in,
height=6.647in,
scale=0.21,
at={(1.85in,0.746in)},
scale only axis,
separate axis lines,
every outer x axis line/.append style={black},
every x tick label/.append style={font=\color{black}},
xmin=0,
xmax=100,
xlabel={Total number of episodes},
every outer y axis line/.append style={black},
every y tick label/.append style={font=\color{black}},
ymin=0,
ymax=70,
ylabel={Number of State Visits},
axis background/.style={fill=white}
]
\addplot [color=red,dash pattern=on 5pt off 5pt]
  table[row sep=crcr]{%
1	1\\
2	1\\
3	1\\
4	1\\
5	2\\
6	3\\
7	4\\
8	4\\
9	4\\
10	4\\
11	4\\
12	5\\
13	6\\
14	6\\
15	6\\
16	7\\
17	7\\
18	7\\
19	7\\
20	8\\
21	9\\
22	10\\
23	10\\
24	10\\
25	11\\
26	11\\
27	12\\
28	13\\
29	13\\
30	14\\
31	14\\
32	14\\
33	15\\
34	15\\
35	15\\
36	15\\
37	15\\
38	15\\
39	15\\
40	16\\
41	16\\
42	16\\
43	17\\
44	18\\
45	18\\
46	18\\
47	19\\
48	19\\
49	19\\
50	19\\
51	19\\
52	20\\
53	21\\
54	22\\
55	23\\
56	23\\
57	23\\
58	23\\
59	23\\
60	24\\
61	25\\
62	25\\
63	26\\
64	26\\
65	27\\
66	27\\
67	27\\
68	27\\
69	27\\
70	27\\
71	28\\
72	29\\
73	29\\
74	30\\
75	31\\
76	31\\
77	32\\
78	33\\
79	33\\
80	33\\
81	33\\
82	33\\
83	34\\
84	34\\
85	34\\
86	34\\
87	34\\
88	35\\
89	35\\
90	36\\
91	36\\
92	36\\
93	36\\
94	36\\
95	36\\
96	36\\
97	37\\
98	37\\
99	38\\
100	38\\
};
\addplot [color=red,width=1.0,solid]
  table[row sep=crcr]{%
1	0\\
2	1\\
3	2\\
4	3\\
5	3\\
6	3\\
7	3\\
8	4\\
9	5\\
10	6\\
11	7\\
12	7\\
13	7\\
14	8\\
15	9\\
16	9\\
17	10\\
18	11\\
19	12\\
20	12\\
21	12\\
22	12\\
23	13\\
24	14\\
25	14\\
26	15\\
27	15\\
28	15\\
29	16\\
30	16\\
31	17\\
32	18\\
33	18\\
34	19\\
35	20\\
36	21\\
37	22\\
38	23\\
39	24\\
40	24\\
41	25\\
42	26\\
43	26\\
44	26\\
45	27\\
46	28\\
47	28\\
48	29\\
49	30\\
50	31\\
51	32\\
52	32\\
53	32\\
54	32\\
55	32\\
56	33\\
57	34\\
58	35\\
59	36\\
60	36\\
61	36\\
62	37\\
63	37\\
64	38\\
65	38\\
66	39\\
67	40\\
68	41\\
69	42\\
70	43\\
71	43\\
72	43\\
73	44\\
74	44\\
75	44\\
76	45\\
77	45\\
78	45\\
79	46\\
80	47\\
81	48\\
82	49\\
83	49\\
84	50\\
85	51\\
86	52\\
87	53\\
88	53\\
89	54\\
90	54\\
91	55\\
92	56\\
93	57\\
94	58\\
95	59\\
96	60\\
97	60\\
98	61\\
99	61\\
100	62\\
};

\addplot [color=blue,dash pattern=on 5pt off 5pt]
  table[row sep=crcr]{%
1	1\\
2	1\\
3	2\\
4	2\\
5	2\\
6	3\\
7	4\\
8	4\\
9	5\\
10	6\\
11	7\\
12	8\\
13	9\\
14	9\\
15	9\\
16	10\\
17	10\\
18	11\\
19	12\\
20	12\\
21	12\\
22	13\\
23	13\\
24	14\\
25	14\\
26	15\\
27	15\\
28	16\\
29	17\\
30	17\\
31	18\\
32	18\\
33	19\\
34	19\\
35	20\\
36	20\\
37	21\\
38	21\\
39	21\\
40	21\\
41	22\\
42	22\\
43	23\\
44	23\\
45	23\\
46	24\\
47	24\\
48	25\\
49	25\\
50	26\\
51	26\\
52	26\\
53	27\\
54	27\\
55	28\\
56	29\\
57	30\\
58	31\\
59	32\\
60	32\\
61	33\\
62	34\\
63	35\\
64	36\\
65	37\\
66	38\\
67	39\\
68	40\\
69	41\\
70	42\\
71	43\\
72	43\\
73	44\\
74	44\\
75	44\\
76	45\\
77	46\\
78	47\\
79	47\\
80	48\\
81	49\\
82	50\\
83	50\\
84	51\\
85	52\\
86	52\\
87	53\\
88	53\\
89	54\\
90	54\\
91	55\\
92	56\\
93	57\\
94	58\\
95	58\\
96	58\\
97	58\\
98	59\\
99	60\\
100	60\\
};
\addplot [color=blue,width=1.0,solid]
  table[row sep=crcr]{%
1	0\\
2	1\\
3	1\\
4	2\\
5	3\\
6	3\\
7	3\\
8	4\\
9	4\\
10	4\\
11	4\\
12	4\\
13	4\\
14	5\\
15	6\\
16	6\\
17	7\\
18	7\\
19	7\\
20	8\\
21	9\\
22	9\\
23	10\\
24	10\\
25	11\\
26	11\\
27	12\\
28	12\\
29	12\\
30	13\\
31	13\\
32	14\\
33	14\\
34	15\\
35	15\\
36	16\\
37	16\\
38	17\\
39	18\\
40	19\\
41	19\\
42	20\\
43	20\\
44	21\\
45	22\\
46	22\\
47	23\\
48	23\\
49	24\\
50	24\\
51	25\\
52	26\\
53	26\\
54	27\\
55	27\\
56	27\\
57	27\\
58	27\\
59	27\\
60	28\\
61	28\\
62	28\\
63	28\\
64	28\\
65	28\\
66	28\\
67	28\\
68	28\\
69	28\\
70	28\\
71	28\\
72	29\\
73	29\\
74	30\\
75	31\\
76	31\\
77	31\\
78	31\\
79	32\\
80	32\\
81	32\\
82	32\\
83	33\\
84	33\\
85	33\\
86	34\\
87	34\\
88	35\\
89	35\\
90	36\\
91	36\\
92	36\\
93	36\\
94	36\\
95	37\\
96	38\\
97	39\\
98	39\\
99	39\\
100	40\\
};

\addplot [color=mycolor1,dash pattern=on 5pt off 5pt]
  table[row sep=crcr]{%
1	0\\
2	1\\
3	2\\
4	3\\
5	3\\
6	3\\
7	3\\
8	3\\
9	3\\
10	4\\
11	5\\
12	6\\
13	7\\
14	8\\
15	8\\
16	8\\
17	8\\
18	8\\
19	9\\
20	10\\
21	10\\
22	10\\
23	10\\
24	11\\
25	12\\
26	12\\
27	13\\
28	14\\
29	14\\
30	15\\
31	16\\
32	17\\
33	18\\
34	19\\
35	20\\
36	20\\
37	20\\
38	21\\
39	21\\
40	21\\
41	22\\
42	23\\
43	24\\
44	25\\
45	25\\
46	26\\
47	27\\
48	28\\
49	28\\
50	29\\
51	30\\
52	31\\
53	32\\
54	33\\
55	34\\
56	35\\
57	36\\
58	37\\
59	37\\
60	37\\
61	38\\
62	39\\
63	40\\
64	41\\
65	42\\
66	43\\
67	43\\
68	43\\
69	44\\
70	45\\
71	45\\
72	46\\
73	47\\
74	48\\
75	48\\
76	48\\
77	49\\
78	50\\
79	51\\
80	51\\
81	52\\
82	53\\
83	54\\
84	54\\
85	55\\
86	56\\
87	56\\
88	57\\
89	57\\
90	58\\
91	59\\
92	60\\
93	61\\
94	62\\
95	63\\
96	63\\
97	63\\
98	64\\
99	65\\
100	66\\
};
\addplot [color=mycolor1,width=1.0,solid]
  table[row sep=crcr]{%
1	1\\
2	1\\
3	1\\
4	1\\
5	2\\
6	3\\
7	4\\
8	5\\
9	6\\
10	6\\
11	6\\
12	6\\
13	6\\
14	6\\
15	7\\
16	8\\
17	9\\
18	10\\
19	10\\
20	10\\
21	11\\
22	12\\
23	13\\
24	13\\
25	13\\
26	14\\
27	14\\
28	14\\
29	15\\
30	15\\
31	15\\
32	15\\
33	15\\
34	15\\
35	15\\
36	16\\
37	17\\
38	17\\
39	18\\
40	19\\
41	19\\
42	19\\
43	19\\
44	19\\
45	20\\
46	20\\
47	20\\
48	20\\
49	21\\
50	21\\
51	21\\
52	21\\
53	21\\
54	21\\
55	21\\
56	21\\
57	21\\
58	21\\
59	22\\
60	23\\
61	23\\
62	23\\
63	23\\
64	23\\
65	23\\
66	23\\
67	24\\
68	25\\
69	25\\
70	25\\
71	26\\
72	26\\
73	26\\
74	26\\
75	27\\
76	28\\
77	28\\
78	28\\
79	28\\
80	29\\
81	29\\
82	29\\
83	29\\
84	30\\
85	30\\
86	30\\
87	31\\
88	31\\
89	32\\
90	32\\
91	32\\
92	32\\
93	32\\
94	32\\
95	32\\
96	33\\
97	34\\
98	34\\
99	34\\
100	34\\
};

\end{axis}
\end{tikzpicture}%
\label{fig:AC_Learn_2} }}%
    \caption{Admission Control training. At the onset, the cumulative rewards are closer together and decisions are more equi-probable. As the agent makes more decisions over subsequent episodes, the admission policy for each of the states becomes more distinct. Weighted fair exploration favors the actions with higher rewards.}%
    \label{fig:AC_Learn}%
\end{figure}

Fig.~\ref{fig:AC_Learn} shows how the learning algorithm for the admission control trades off exploration and exploitation using our weighted fair exploration scheme; cf. Eq.~\eqref{eq:selectioner}. The evolution of the accumulated reward for a subset of 3 state-action pairs is shown in Fig.~\ref{fig:AC_Learn_1}. In the initial learning phases, the difference between the values is not as distinct and a DROP or ADMIT decision is made with about the same probability. However as the admission control continues to update the accumulated reward values at every episode, the comparative value in each state-action pair becomes more distinct. The latter results in more ADMIT than DROP decisions in the regions spanning the lower utilization levels, and the converse is true in the regions spanning the high utilization levels as shown in Fig.~\ref{fig:AC_Learn_2}.

At convergence the admission policy for each region is fully determined. With reference to Fig.~\ref{fig:AC_Learn_1}, strong ADMIT and strong DROP policies are at the extreme levels of low (dotted and solid red lines) and high (dotted and solid teal lines) utilization respectively. In the mid levels of utilization (dotted and solid teal lines), the difference in the decision variable~(\textit{Q}-Value) is not as stark, resulting in a weak DROP decision at the boundary.

The limiting value of utilization $x_\mathrm{lim}$, beyond which service times are generally unpredictable, is learned to be 45\%. The admission controller therefore drops requests sent to a selected VM whose utilization is higher than this value. We recall that the load balancer selects the VM with the lowest, most recently logged utilization value at the time of receiving a request.
\subsection{Horizontal Scaling}
The horizontal scaling profile produced by the scheme in~\cite{Gandhi} in response to the test traffic~(cf. Fig.~\ref{fig:Offered_Load}) is shown in Fig.~\ref{fig:Scale_EKF}. The target response time used was $R_{\mathrm{sla}}=5~\mathrm{\mu s}$ per job. The scaling behaviour in this scheme is quite stiff, given its reliance on the EKF, which may filter out bursty traffic that require greater agility.

The scaling profile obtained from our proposed SQLR scheme, in reference to the test traffic, is shown in Fig.~\ref{fig:SQLR_Scaling}. The behaviour of the scaler steadily improves with increased exposure to the test traffic. As more states converge, the scaling behaviour becomes more predictable as seen by moving from Fig.~\ref{fig:SQLR_A} to Fig.~\ref{fig:SQLR_Sim} and Fig.~\ref{fig:SQLR_Case1_2}. The number of VMs provisioned oscillates around a suitable number that best achieves the compromise of resource cost and penalties encapsulated in the training parameters $\theta$ and $\beta$.

Moreover, in Fig.~\ref{fig:SQLR_Sim}, we see that the first intervals to exhibit convergence (implied by greater stability in the scaling behaviour) are those with higher similarity to the training traffic~(cf. Fig.~\ref{fig:Training_Load}). For instance the intervals of hours 6 to 8 and 16 to 18 with an average of 40 requests per minute (cf. Fig.~\ref{fig:Offered_Load}), closely resemble those of hours 1 to 4 and 20 to 23 of the training traffic (cf. Fig.~\ref{fig:Training_Load}). SQLR thus has the ability to re-use contextual knowledge learned from one workload on a subsequent one with similar characteristics.

Assigning different values to the training parameters $\theta$ and $\beta$ results in different policies being learned by SQLR. As shown in Fig.~\ref{fig:SQLR_Case1_2}, a low value of $\theta$ relative to $\beta$ (Case~1) results in cost-focused scaling policies that emphasize resource cost more than service unavailability due to blocking. When $\theta \gg \beta$, as in Case~2, more service-focused policies are learned giving greater importance to service availability than to resource cost.
\begin{figure}[t]
    \centering
    \begin{tikzpicture}
\begin{axis}[%
width=12.917in,
height=6.647in,
scale=0.21,
at={(2.167in,0.897in)},
scale only axis,
separate axis lines,
every outer x axis line/.append style={black},
every x tick label/.append style={font=\color{black}},
xmin=0,
xmax=24,
xtick={0,6,12,18,24},
xticklabels={0,6,12,18,24},
xlabel={Ref. Time(Hr)},
every outer y axis line/.append style={black},
every y tick label/.append style={font=\color{black}},
ymin=0,
ymax=5,
ylabel={Number of VMs},
axis background/.style={fill=white}
]
\addplot [color=red,solid,forget plot]
  table[row sep=crcr]{%
0	1\\
4.05875555555556	1\\
4.05875555555556	2\\
4.48516527777778	2\\
4.48516527777778	1\\
4.53528972222222	1\\
4.53528972222222	2\\
4.786175	2\\
4.786175	1\\
4.9615675	1\\
4.9615675	2\\
5.48825777777778	2\\
5.48825777777778	1\\
5.538385	1\\
5.538385	2\\
5.63894083333333	2\\
5.63894083333333	1\\
5.73915083333333	1\\
5.73915083333333	2\\
8.0708325	2\\
8.0708325	3\\
8.54774194444444	3\\
8.54774194444444	2\\
8.59791944444444	2\\
8.59791944444444	3\\
8.67343055555556	3\\
8.67343055555556	2\\
9.07457	2\\
9.07457	3\\
9.45120416666667	3\\
9.45120416666667	2\\
9.501375	2\\
9.501375	3\\
9.55182277777778	3\\
9.55182277777778	2\\
9.60196805555556	2\\
9.60196805555556	3\\
9.80294611111111	3\\
9.80294611111111	2\\
9.85311	2\\
9.85311	3\\
11.1833677777778	3\\
11.1833677777778	4\\
11.2840769444444	4\\
11.2840769444444	3\\
11.4346905555556	3\\
11.4346905555556	4\\
11.5353997222222	4\\
11.5353997222222	3\\
11.7111416666667	3\\
11.7111416666667	4\\
12.5651194444444	4\\
12.5651194444444	3\\
12.6153077777778	3\\
12.6153077777778	4\\
13.0424333333333	4\\
13.0424333333333	3\\
15.1252372222222	3\\
15.1252372222222	2\\
15.2004641666667	2\\
15.2004641666667	3\\
15.3011094444444	3\\
15.3011094444444	2\\
15.3512558333333	2\\
15.3512558333333	3\\
15.4518733333333	3\\
15.4518733333333	2\\
15.5020436111111	2\\
15.5020436111111	3\\
15.6277980555556	3\\
15.6277980555556	2\\
15.7030425	2\\
15.7030425	3\\
16.0294741666667	3\\
16.0294741666667	2\\
18.1108502777778	2\\
18.1108502777778	1\\
18.1609913888889	1\\
18.1609913888889	2\\
18.2615133333333	2\\
18.2615133333333	1\\
18.3367119444444	1\\
18.3367119444444	2\\
18.41222	2\\
18.41222	1\\
18.5625744444444	1\\
18.5625744444444	2\\
18.7884919444444	2\\
18.7884919444444	1\\
18.8386563888889	1\\
18.8386563888889	2\\
18.8891019444444	2\\
18.8891019444444	1\\
18.9392141666667	1\\
18.9392141666667	2\\
19.2905922222222	2\\
19.2905922222222	1\\
19.3407316666667	1\\
19.3407316666667	2\\
19.3911377777778	2\\
19.3911377777778	1\\
19.5164655555556	1\\
19.5164655555556	2\\
20.0432355555556	2\\
20.0432355555556	1\\
24	1\\
};
\end{axis}
\end{tikzpicture}%
    \caption{VM Scaling for the EKF-based horizontal scaling scheme proposed in~\cite{Gandhi}. }
    \label{fig:Scale_EKF}    
\end{figure}
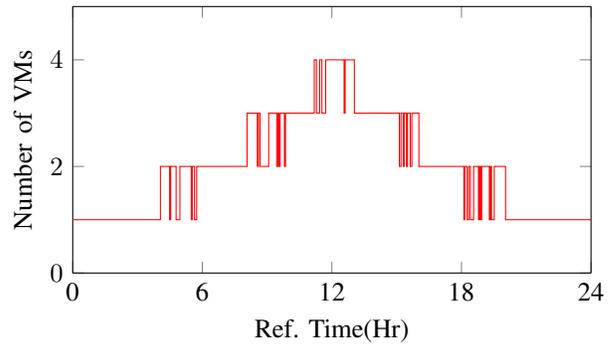
\begin{figure}[h]%
    \centering
    \subfloat[At approx. 10\% convergence. With $\theta=1.0$, $\beta=0.01$.]{{\input{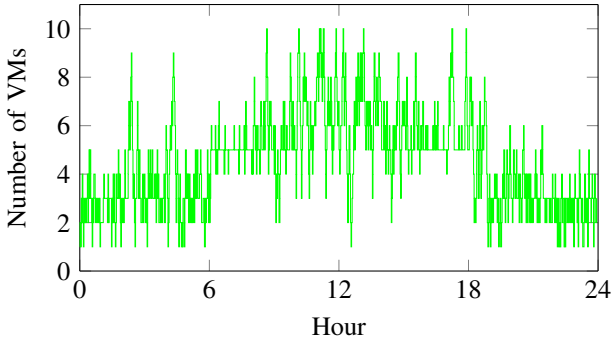}\label{fig:SQLR_A} }}%
    \qquad
    \subfloat[At approx. 20\% convergence. With $\theta=1.0$, $\beta=0.01$. Early stabilization  from 6-8 and 16-18 is due to the similarity of the workload experienced in these regions with that of the training traffic profile.]{{\input{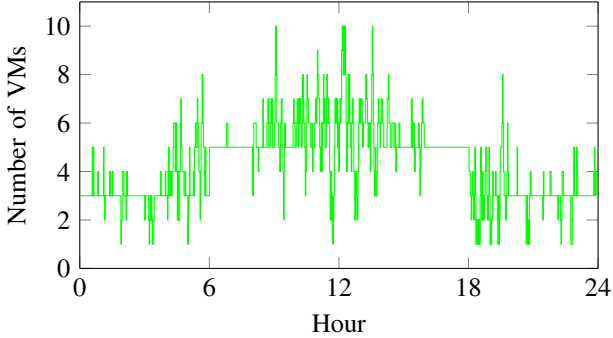}\label{fig:SQLR_Sim} }}%
    \qquad
    \subfloat[At approx. 50\% convergence.]{{\input{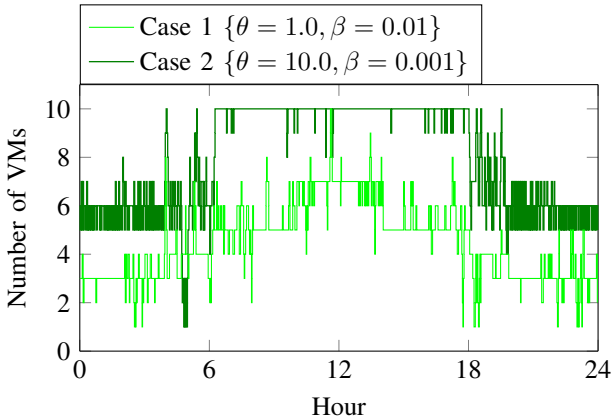}\label{fig:SQLR_Case1_2} }}%
    \caption{SQLR scaling behaviour with experience after pre-training with $P_{\mathrm{blk}}=0.001$. The convergence level is the proportion of unique states visited for which $\epsilon=\epsilon_{\mathrm{min}}$.}%
    \label{fig:SQLR_Scaling}%
\end{figure}
\begin{figure}
    \centering
    \input{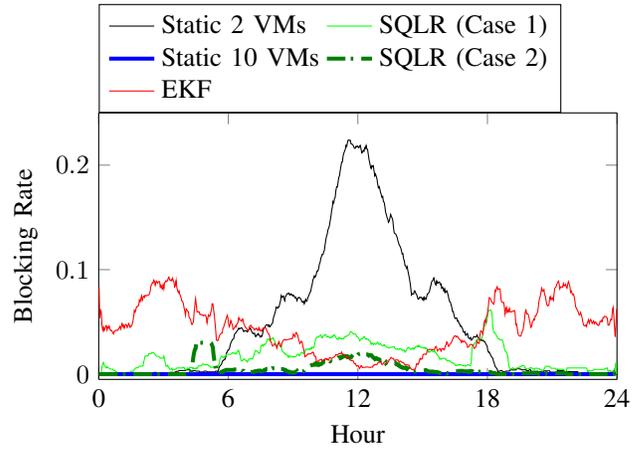}
    \caption{Blocking rates over two-minute intervals. Two SQLR configurations are shown: Case~1~($\theta=1.0,  \beta=0.01$) and Case~2~($\theta=10.0, \beta=0.001$). For clarity, a moving average filter is applied with a window size of 30 samples.}
    \label{fig:Pb_Time}
\end{figure}
\begin{figure}
    \centering
    \input{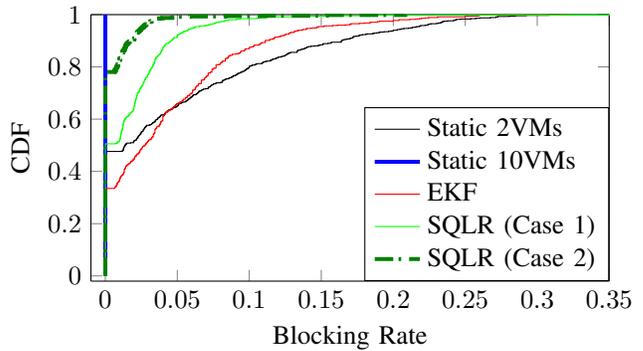}
    \caption{Blocking rate distribution. Two SQLR configurations are shown: Case~1($\theta=1.0, \beta=0.01$) and Case~2($\theta=10.0, \beta=0.001$).}
    \label{fig:Pb_Dist}
\end{figure}
\subsection{Blocking Rates}
The exploration mechanism of the \textit{Q}-Learning algorithm at the core of SQLR means that it may sometimes make sub-optimal decisions in less known states, resulting in under-provisioning (such as at hour 18 for case~1, and at hour 5 for case~2 in  Fig.~\ref{fig:SQLR_Case1_2}). This results in relatively high blocking rates as shown in Fig.~\ref{fig:Pb_Time}. Our guided fair exploration mechanism ameliorates the effects of such under-provisioning ensuring that their duration is short.

Since the EKF scaler relies on workload measurements to predict response times and scale accordingly, it is particularly susceptible to under-estimating resource requirements when demand is low. This is evident at off-peak intervals in Fig.~\ref{fig:Pb_Time} where, between hours 0 and 7 and between hours 17 and 24 it averages an allocation of 1 VM resulting in considerable blocking much higher than the other schemes.

Static provisioning results in significant under or over-provisioning, as exhibited by the 2 VMs and 10 VMs curves respectively. This situation is clearly untenable given that serious penalties are levied on the CSP for service unavailability on the one hand; and on the other hand, significant yet unnecessary operational expenditure is incurred to maintain superfluous resources.

Fig.~\ref{fig:Pb_Dist} compares the distribution of blocking rates for the provisioning mechanisms. We take the static over-provisioned case (with 10 VMs) as a reference benchmark with 0\% blocking and resources in terms of VM-hours. With reference to this benchmark, SQLR Case~2 saves up to 25\% of resources with less than 5\% blocking for 99\% of the requests. SQLR Case~1 saves up to 55\% of resources with less than 5\% blocking for 92\% of the requests. The EKF-based scaler saves almost 80\% VM-Hours but at the expense of service availability, only 65\% of the requests are served with less than 5\% of them being blocked. The static under-provisioned case of 2 VMs achieves similar savings as the EKF-based scaler with similar performance at the 5\% blocking reference. However, when we consider requests with up to 10\% blocking, the EKF-based scaler only achieves this for 80\% of the requests but static provisioning with 2 VMs achieves this for 88\% of the requests.
\subsection{Service times}
\begin{figure}[t]%
    \centering    
    \includegraphics{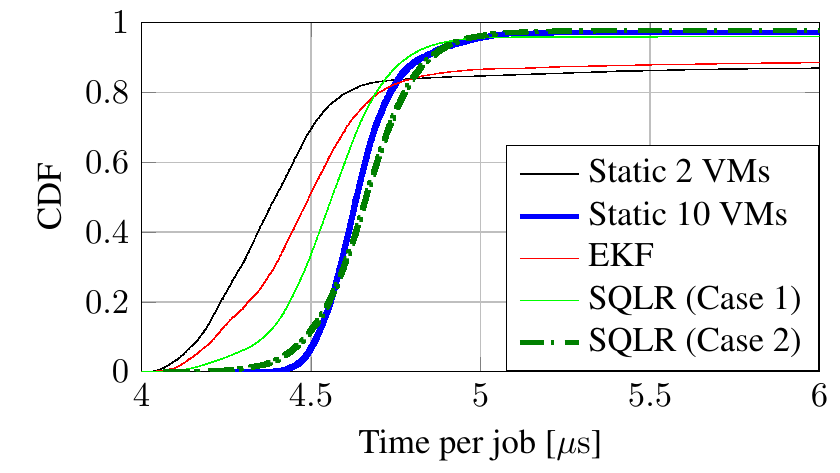}
    \caption{Service time distribution per job. Two SQLR configurations are shown: Case~1($\theta=1.0, \beta=0.01$) and Case~2($\theta=10.0, \beta=0.001$). The service time for each request is divided by the corresponding number of iterations it generates to obtain the time per job.}
    \label{fig:ST_Dist}
\end{figure}

The distribution of service times is shown in Fig.~\ref{fig:ST_Dist}. We obtain the service time per job by dividing the service time of each request by the corresponding number of proof-of-work iterations it generates. These response times include the administrative overhead incurred by the hypervisor to switch between the host and guest while managing virtual machines. It also includes context switching between user mode and kernel mode of the corresponding operating systems.

This overhead increases with the number of virtual machines being administered as well as with how often they switch contexts. Dynamic scaling, which entails starting up and shutting down VMs, exacerbates the latter. Owing to the combination of these factors, SQLR case~2 (with greater penalties for blocking than resource usage) is impacted greatly by this phenomenon. This is because its policies implicitly employ more VMs. It closely follows the static over-provisioned case with 10 VMs, that incurs high administrative overhead throughout. 

However, the over-provisioned scenario still provides the ideal case with the lowest-variance (highly predictable) service times. Both SQLR cases closely approach this ideal case with about 96\% of the requests being served within $5~\mathrm{\mu s}$ per job compared to 94\% for the over-provisioned case and only 87\% for the EKF case. 

Moreover, for SQLR case~2 with $\theta=10.0$ and $\beta=0.001$, the improvement in the proportion of responses within the cutoff service time of $5~\mathrm{\mu s}$ is only  marginal compared to the more cost-focused case~1 with $\theta=1.0$ and $\beta=0.01$ as shown in Fig.~\ref{fig:SQLR_Case1_2}. This is despite the extra amount of resources that case~2 deploys compared to case~1. This represents a diminishing return on policies biased to provision more VMs owing primarily to the additional administrative overheads incurred as a consequence.
\begin{figure}%
    \centering
    \input{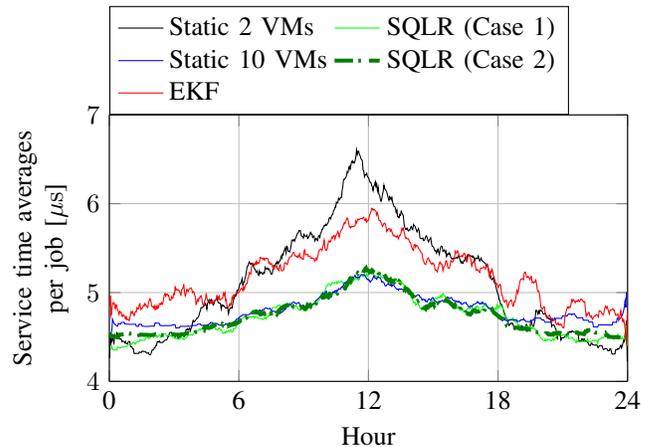}
    \caption{Moving averages of service times (taken over a window of 30 samples to smooth out switching overheads). Two SQLR configurations are shown: Case~1($\theta=1.0, \beta=0.01$) and Case~2($\theta=10.0, \beta=0.001$).}
    \label{fig:Combined_ST}
\end{figure}
\begin{figure*}[p]
    \centering
    \subfloat[Frequency of blocking. The number of responses exceeding $5~\mathrm{\mu s}$ as a fraction of the total number of responses. The white region indicates unexplored resource allocations for the offered load.]{\scalebox{0.8}{\input{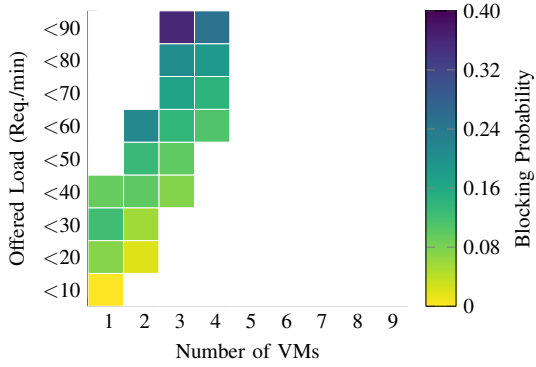} }\label{fig:Pb_Heat_EKF_A}}%
    \qquad
    \subfloat[Severity of blocking. The mean deviation from $5~\mathrm{\mu s}$ of the responses exceeding $5~\mathrm{\mu s}$. The white region includes responses within the service time limit as well as unexplored allocations.]{\scalebox{0.8}{\input{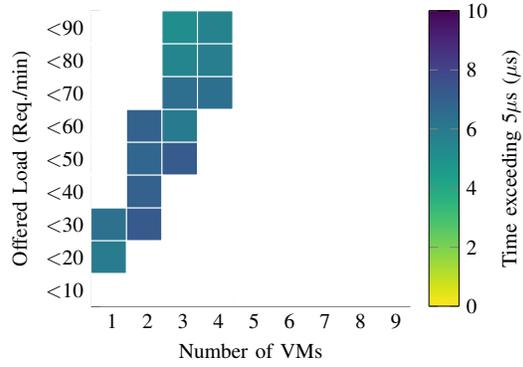} }\label{fig:Pb_Heat_EKF_B}}%
    \caption{Soft Blocking Probability for EKF Scaler. 87\% of the responses are within $5~\mathrm{\mu s}$ per job.}%
    \label{fig:Pb_Heat_EKF}%
\end{figure*}%
\vspace{0.2em}%
\begin{figure*}%
    \centering
    \subfloat[Frequency of blocking. The number of responses exceeding $5~\mathrm{\mu s}$ as a fraction of the total number of responses. The white region indicates unexplored resource allocations for the offered load.]{\scalebox{0.8}{\input{Blocking_Heat_SQLR_1.tex} } \label{fig:Pb_Heat_SQLR_A}}%
    \qquad
    \subfloat[Severity of blocking. The mean deviation from $5~\mathrm{\mu s}$ of the responses exceeding $5~\mathrm{\mu s}$. The white region includes responses within the service time limit as well as unexplored allocations.]{\scalebox{0.8}{\input{Blocking_Nuance_SQLR_1.tex} }\label{fig:Pb_Heat_SQLR_B}}%
    \caption{Soft Blocking Probability for SQLR Case~1 ($\theta=1.0$, $\beta=0.01$). 95.6\% of the responses are within $5~\mathrm{\mu s}$ per job. The instances of high severity are mainly due to exploratory scale-in actions in states where $\epsilon>\epsilon_{\mathrm{min}}$.}%
    \label{fig:Pb_Heat_SQLR}%
\end{figure*}%
\vspace{0.2em}%
\begin{figure*}%
    \centering
    \subfloat[Frequency of blocking. The number of responses exceeding $5~\mathrm{\mu s}$ as a fraction of the total number of responses. The white region indicates unexplored resource allocations for the offered load.]{\scalebox{0.8}{\input{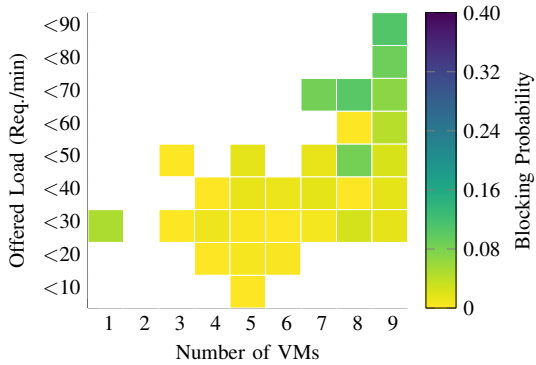} }\label{fig:Pb_Heat_SQLR2_A}}%
    \qquad
    \subfloat[Severity of blocking. The mean deviation from 5$\mu$s of the responses exceeding 5$\mu$s. The white region includes responses within the service time limit as well as unexplored allocations.]{\scalebox{0.8}{\input{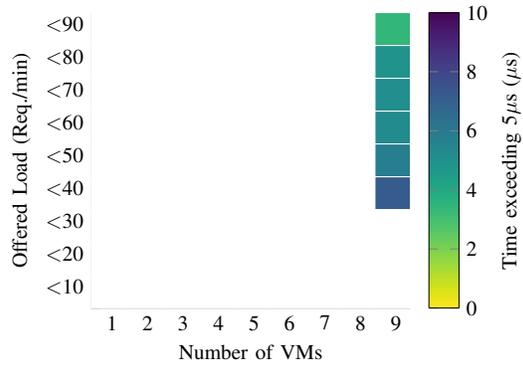} }\label{fig:Pb_Heat_SQLR2_B}}%
    \caption{Soft Blocking Probability for SQLR Case~2~($\theta=10.0$, $\beta=0.001$). 96.2\% of the responses are within $5~\mathrm{\mu s}$ per job. The instances of high severity are mainly due to exploratory scale-in actions in states where $\epsilon>\epsilon_{\mathrm{min}}$.}%
    \label{fig:Pb_Heat_SQLR2}%
\end{figure*}

In order to compare the scaling schemes without the biasing effect of the administrative overhead, we average out the effect. This is akin to noise filtering in communication systems. We do this by first obtaining the average service times over two-minute intervals, and then applying a moving average filter having a window of 30 samples. Since context switching happens in the order of clock cycles, the two above operations over intervals that are orders of magnitude longer than a clock cycle nearly nullify the switching overhead.

When we apply the operations stated above to the service time per job, we obtain the results depicted in Fig.~\ref{fig:Combined_ST}. Both SQLR cases considered, closely follow the over-provisioned one with the ideal response times. At low traffic (hours 0-7 and 17-24), the administrative costs for maintaining a large number of VMs outweigh the gains made in improving service times by using more resources. Over these intervals, SQLR performs slightly better than the unconstrained case by provisioning fewer VMs. Conversely, the EKF-based scaler still under-performs, since the single VM it provisions over these intervals is not sufficient to meet the demand within the cut-off service time.

This marked difference in response times, owing to differences in the scaling mechanisms, is clearly depicted in Figs.~\ref{fig:Pb_Heat_EKF} to~\ref{fig:Pb_Heat_SQLR2} where we compare the soft-blocking performance (defined as the proportion of admitted requests whose service times extend beyond our cut-off of $5~\mathrm{\mu s}$ per job). In these heatmaps, only blocks with statistical significance (30 or more responses) are considered. Moreover, the white region indicates unexplored resource allocations for the offered load. Additionally, for the severity heatmaps in Figs.~\ref{fig:Pb_Heat_EKF_B}, \ref{fig:Pb_Heat_SQLR_B} and \ref{fig:Pb_Heat_SQLR2_B}, the white region indicates those allocations leading to service times within the limit of $5~\mathrm{\mu s}$. Moreover, the offered load values on the y-axis indicate the upper bound with the value immediately below indicating the lower bound i.e. \textquotedblleft\textless20\textquotedblright\ indicates the interval [10,20) requests per min.

As depicted in Fig.~\ref{fig:Pb_Heat_EKF}, the EKF scaler employed by~\cite{Gandhi} is prone to overruns even under light loads given that it is very conservative in allocating extra VMs. As a result, this increases the strain on the few that are assigned. In our scheme, whose responses are shown in Figs.~\ref{fig:Pb_Heat_SQLR} and~\ref{fig:Pb_Heat_SQLR2}, a smaller proportion of service times exceed the cut-off time (particularly at moderate to high offered loads). The reason is that SQLR is more sensitive to abrupt changes in traffic and assigns resources in a more agile fashion compared to the EKF-based scheme. In fact, the effect of the filter on workload measurements when making scaling decisions makes the latter scheme oblivious to short-lived bursts in demand.

Further, comparing the SQLR cases shown in Figs.~\ref{fig:Pb_Heat_SQLR} and \ref{fig:Pb_Heat_SQLR2}, the provisioning policies of Case~2 result in fewer instances of soft blocking than Case~1. This is because the latter provisions fewer VMs which increases the likelihood of operating them at higher CPU loads hence longer service times as alluded to in Eq.~\eqref{eq:QService} and shown in Fig.~\ref{fig:Resp_CPU}. The cases of high severity (particularly in Case~2) are because of exploratory actions at high demand whereby SQLR momentarily scales in. However, by evaluating the sub-optimality of these actions, our weighted fair guided exploration quickly scales out as is evident around hours 10 and 12 in Fig.~\ref{fig:SQLR_Case1_2}.
\section{Conclusions and Future Work}
\label{FinisTerrae}
We have presented an agile horizontal scaling system, SQLR, that learns the most appropriate horizontal scaling decision to take (without any fore-knowledge of the underlying system configuration) under highly dynamic workloads. We show that our modified \textit{Q}-Learning scheme enables our system to learn multiple policies and re-use any applicable knowledge to new traffic profiles exhibiting previously encountered characteristics.

SQLR progressively self-optimizes resource cost and service availability constraints to achieve the proper trade-off. These constraints can be supplied by the CSP as external input after proper determination from their business processes. Such high-level objectives make SQLR easily configurable and adaptable to any cloud application as no domain-specific knowledge is required. We contrast this with a state-of-the-art scaling system and show that our scheme achieves better performance, similar to that of an over-provisioned system.

As with most machine learning-based systems, our scheme is subject to a training overhead. However, because of its capacity for contextual knowledge re-use, it can be trained offline with representative workloads. Also given our weighted fair exploration mechanism, any subsequent residual learning can be done in production workloads with a much reduced risk of poor decisions in the process. We show that with only about 50\% of fully converged states, our scheme performs almost as well as the unconstrained resource benchmark~(static over-provisioning).

In the future, we intend to apply the short-term memory \textit{Q}-Learning employed in this work to migrate VMs to other hosts when the physical resources of the active host are under strain. We also plan to use a variant of the scheme to carry out orchestration of distributed cloud applications.

\enlargethispage{-10cm}

\begin{IEEEbiographynophoto}{Constantine Ayimba}
received his M.Sc. in Wireless Communications from Lund University (Sweden) in 2016 where he was a Swedish Institute Scholar. He previously held positions in Ericsson and Strathmore University (Kenya). He obtained his BSc. in Electrical Engineering from the University of Nairobi (Kenya) in 2006. He is currently pursuing his Ph.D. at IMDEA Networks Institute (Spain) where his research focuses on machine learning for self-driving networks.
\end{IEEEbiographynophoto}
\begin{IEEEbiographynophoto}{Paolo Casari}
received the PhD in Information Engineering in 2008 from the University of Padova, Italy. He was on leave at the Massachusetts Institute of Technology in 2007, working on underwater communications and networks. He collaborated to several funded projects including CLAM (FP7), RACUN (EDA), as well as US ARO, ONR and NSF initiatives. He is the PI of the NATO SPS project ThreatDetect, and the scientific coordinator of the EU H2020 RECAP and SYMBIOSIS projects. In 2015, he joined the IMDEA Networks Institute, Madrid, Spain, where he led the Ubiquitous Wireless Networks group. As of October 2019, he joined the faculty of the University of Trento, Italy. He regularly serves in the organizing committee of several international conferences, and is currently on the editorial boards of the IEEE Transactions on Mobile Computing and of the IEEE Transactions on Wireless Communications. Previously, he has been guest editor of a special issue of IEEE Access on ``Underwater Acoustic Communications and Networking,'' as well as of a special issue of the Hindawi Journal of Electrical and Computer Engineering on the same topic. He received two best paper awards. His research interests include diverse aspects of networked communications, such as channel modeling, network protocol design, localization, simulation, and experimental evaluations.
\end{IEEEbiographynophoto}
\begin{IEEEbiographynophoto}{Vincenzo Mancuso}
is Research Associate Professor at IMDEA Networks Institute, Madrid, Spain, and recipient of a Ramon y Cajal research grant of the Spanish Ministry of Science and Innovation. Previously, he was with INRIA Sophia Antipolis (France), Rice University (Houston, TX, USA) and University of Palermo (Italy), from where he obtained his MSc and his Ph.D. in Electronics, Computer
Science and Telecommunications. He has authored more than 110 peer-reviewed publications focusing on Internet QoS and on the analysis, design, and experimental evaluation of opportunistic and adaptive protocols and architectures for wireless networks. He is currently working on analysis and optimization of wireless access networks and on the measurements and assessment of mobile broadband networks.
\end{IEEEbiographynophoto}
\vfill
\end{document}